\documentclass[aps,preprint,nofootinbib,showpacs,superscriptaddress]{revtex4}
\usepackage{amssymb}
\usepackage{amsmath}
\usepackage{amsfonts}
\usepackage{graphicx}
\usepackage{multirow}
\usepackage{subfigure}
\usepackage{float}
\usepackage{graphics}
\usepackage{slashed}
\usepackage{color}
\usepackage[hypertex]{hyperref}

\def\nslash{n\!\!\!\slash}
\def\bnslash{\bar n\!\!\!\slash}

\newcommand{\nn}{\nonumber}

\newcommand{\bn}{\bar n}

\newcommand{\mcdot}{\!\cdot\!}

\begin{document}
\title{Transverse momentum resummation for color sextet and antitriplet scalar production at the LHC}
\author{Ze Long Liu}
\affiliation{School of Physics and State Key Laboratory of
Nuclear Physics and Technology, Peking University, Beijing 100871,
China}
\author{Chong Sheng Li}
\email{csli@pku.edu.cn} \affiliation{School of Physics and State
Key Laboratory of Nuclear Physics and Technology, Peking University,
Beijing 100871, China} \affiliation{Center for High Energy Physics,
Peking University, Beijing 100871, China}
\author{Yan Wang}
\affiliation{School of Physics and State Key Laboratory of
Nuclear Physics and Technology, Peking University, Beijing 100871,
China}
\author{Yong Chuan Zhan}
\affiliation{School of Physics and State Key Laboratory of
Nuclear Physics and Technology, Peking University, Beijing 100871,
China}
\author{Hai Tao Li}
\affiliation{School of Physics and State Key Laboratory of
Nuclear Physics and Technology, Peking University, Beijing 100871,
China}

\date{\today}

\begin{abstract}
We study the factorization and resummation of the transverse momentum spectrum of the color sextet and antitriplet scalars produced at the LHC based on soft-collinear effective theory. Compared to $Z$ boson and Higgs production, a soft function is required to account for the soft gluon emission from the final-state colored scalar. The soft function is calculated at the next-to-leading order, and the resummation is performed at the approximate next-to-next-to-leading logarithmic accuracy. The non-perturbative effects and PDF uncertainties are also discussed.

\end{abstract}


\maketitle

\section{INTRODUCTION}\label{s1}
The Large Hadron Collider (LHC) provides a great opportunity to search for new physics beyond the Standard Model (SM). For example, observation of the color sextet (antitriplet) scalars will be a direct signal of new physics beyond the SM. In fact, the color sextet (antitriplet) scalars have been included in many new physics models, such as unification theories~\cite{Pati:1974yy,Mohapatra:1980qe,Perez:2007rm}, Supersymmetry with R-parity violation~\cite{Barbier:2004ez}, diquark Higgs~\cite{Mohapatra:2007af}, et al. So it is preferable to concern with such signal in the model independent way rather than considering some specific models. A colored scalar can be produced in quark-quark fusion with color structure obtained from ${\bf 3}\times{\bf 3}={\bf 6}\oplus{\bf{\bar 3}}$, where ${\bf 3}$, ${\bf 6}$ and ${\bf{\bar 3}}$ are the triplet, sextet and antitriplet representations of the $SU(3)_C$ color group. The interaction of the color sextet (antitriplet) scalars with two quarks can be written as
\begin{equation}\label{eq:lagrangian}
 \mathcal L=2\sqrt2\left[\bar K_i{}^{ab}\phi^i\bar \psi_a\big(\lambda_LP_L+\lambda_RP_R\big)\psi^C_b+\text{h.c.}\right],
\end{equation}
where $P_{L,R}\equiv(1\mp\gamma_5)/2$ are the usual left- and right-hand projectors, $\lambda_{L/R}$ is the Yukawa like coupling, $a$, $b$ are the color indices, and $\bar K_i{}^{ab}$ is Clebsch-Gordan coefficients \cite{Han:2009ya}. $q^C$ is the charge conjugate quark field, and the sum over quark flavors has been suppressed. The scalar $\phi^i$ transforms according to either sextet or antitriplet representation of $SU(3)_C$. The production of a heavy resonance via quark-quark fusion is significantly enhanced at the LHC for larger values of the partonic Bjorken-$x$, because valence quarks have large parton density there, where the gluon density drops off rapidly.

The exotic colored states attract a lot attention in experiments \cite{Chatrchyan:2013qha,Aad:2012em}. The most current data reported by the CMS experiment \cite{Chatrchyan:2013qha} at $\sqrt{s}=8\,{\rm TeV}$ corresponding to an integrated luminosity of 4 $\rm fb^{-1}$ excludes $\rm E_6$ diquarks \cite{Hewett:1988xc} with mass less than 4.28 TeV at 95\% confidence level. As shown in Ref.~\cite{Mohapatra:2007af,Chen:2009xjb,Chen:2008hh}, the measurements of $D^0$-$\overline{D^0}$ mixing and the rate of $D\rightarrow \pi^+\,\pi^0(\pi^+\phi)$ decay can constrain the couplings of the colored scalars to two up-type quarks $\lambda_R^{uu},\lambda_R^{uc} \le 0.1$, $|{\rm Re}(\lambda^{cc}\lambda^{uu*})|\sim 5.76\times 10^{-7}$ for $m_\phi\sim 1\,{\rm TeV}$. In addition, the left-handed coupling $\lambda_L$ also gets tight constraints due to minimal flavor violation.
Since we use the model independent coupling $\lambda^2=\lambda_L^2+\lambda_R^2$, above constraints can be relaxed in the following scenario.

Resonant production of the color antitriplet scalars and vectors has been calculated at the leading order (LO), respectively, in Ref.~\cite{Han:2010rf,Atag:1998xq,Arik:2001bc,Cakir:2005iw}, and pair production of the color sextet scalars has been studied at the LO in Ref.~\cite{Chen:2008hh,Richardson:2011df}. In Ref.~\cite{Han:2009ya},the next-to-leading order (NLO) QCD corrections to the production of color sextet and antitriplet scalars have also been calculated . The decay of color triplet and sextet has also been studied in Ref.~\cite{Gogoladze:2010xd} and Ref.~\cite{Berger:2010fy,Karabacak:2012rn}, respectively.
Very recently, the threshold resummation for the production of a color sextet (antitriplet) has been investigated in Ref.~\cite{Zhan:2013sza}. As is well-known in the case of Drell-Yan and Higgs production, the fixed-order predictions are unreliable in small $q_T$ region, because soft and collinear gluon emissions give rise to large logarithms of scale ratio $\ln(q_T^2/Q^2)$ at each order in perturbation theory, where $Q\gg q_T$ is a typical hard scale of the process. For the case of Drell-Yan and Higgs, the method to deal with this problem is the so-called Collins-Soper-Sterman (CSS) formalism \cite{Collins:1981uk,Collins:1981va,Collins:1984kg}, in which the large logarithms can be resummed to all orders in the strong coupling $\alpha_s$. For colored scalar production, the CSS formulism can not be directly applied due to gluon emissions from the colored scalar in the final state. Nevertheless, Ref.~\cite{Han:2009ya} achieved the transverse momentum resummation for the production of a colored scalar at the leading logarithmic (LL) level by modifying the CSS formulism.

In this paper, we investigate the transverse momentum resummation in single production of the color sextet (antitriplet) scalars  at the LHC with the approximate next-to-next-leading logarithmic ($\rm NNLL_{approx}$) accuracy in the framwork of the soft-collinear effective theory (SCET) \cite{Bauer:2000yr,Bauer:2001yt,Beneke:2002ph}. The framework is built upon the works in Ref.~\cite{Becher:2010tm,Becher:2011xn,Zhu:2012ts,Li:2013mia}, which systematically resum the large logarithms to arbitrary accuracy. A novel feature of the method in the framework of SCET is the appearance of a transverse soft function, which describes color exchange among the initial state and final state particles.

This paper is organized as follows. In Sec.~\ref{sec:factor}, we briefly show the derivation of the factorization formula for the single colored scalar production at the LHC. In Sec.~\ref{sec:resum}, we calculate the hard function and the soft function at the NLO, and show the resummation formula at the $\rm NNLL_{approx}$. We expand the resummation formula to the NLO in Sec.~\ref{sec:qtfo} and compare them with the exact NLO calculation at small transverse momentum. In Sec.~\ref{sec:numdis}, we discuss the scale and PDF uncertainties of the cross section, and compare our numerical results with the ones in Ref.~\cite{Han:2009ya}. We conclude in Sec.~\ref{sec:conclusion}

\section{Derivation of the factorization formular}\label{sec:factor}
In this section we present the derivation of the factorization for the production of a color sextet (antitriple) scalar using SCET. The transverse momentum resummation discussed has some similarity with threshold resummation~\cite{Zhan:2013sza}, for example, the hard function which encodes the short-distance physics is exactly the same as the one in threshold resummation. But it is genuinely different from that, since the treatment of soft and collinear radiations are completely different from the threshold resummation.

We study the production of a colored scalar with mass $m_\phi$ and transverse momentum $q_T$ in the kinematic region where $m_\phi^2\gg q_T^2\gg\Lambda_{\rm QCD}^2$. To describe collinear and soft fields in SCET, it is convenient to define two light-like vectors along the beam directions $n^\mu=(1,0,0,1)$ and $\bn^\mu=(1,0,0,-1)$, which satisfy $n\cdot\bn=2$. We can decompose any four-vector with respect to $n^\mu$ and $\bar{n}^\mu$ as
\begin{equation}\label{eq:lightconedecp}
  p^\mu\,=\,(n\cdot p)\frac{\bn^\mu}{2}\,+\,(\bn\cdot p)\frac{n^\mu}{2}\,+\,p_\perp^\mu
       \,=\,p_+\frac{\bn^\mu}{2}\,+\,p_-\frac{n^\mu}{2}\,+\,p_\perp^\mu\,\,.
\end{equation}
We define a small parameter $\lambda=q_T / m_\phi$ and quote the components $(p_+,p_-,p_\perp)$ of momentum. The relevant momentum regions are
\begin{eqnarray}
   \mbox{hard:} \quad\,\,
    p_{h} &\sim& m_\phi\,(1,1,1) \,, \nonumber\\
   \mbox{hard-collinear:} \quad
    p_{hc} &\sim& m_\phi\,(\lambda^2,1,\lambda) \,,
    \nonumber\\
   \mbox{anti-hard-collinear:} \quad
    p_{\overline{hc}} &\sim& m_\phi\,
    (1,\lambda^2,\lambda) \,, \nonumber\\
   \mbox{soft:} \quad\,\,
    p_s &\sim& m_\phi\,(\lambda,\lambda,\lambda) \,.
    \nonumber
\end{eqnarray}

We consider the process
\begin{equation}\label{eq:preocess}
 N_1(P_1)+N_2(P_2)\rightarrow \phi(q)+X(p_X)\,,
\end{equation}
where $N_1$ and $N_2$ are the incoming hadrons and $X$ are the inclusive hadronic final states. For later convenience, we define the following kinematic variables
\begin{equation}
   s=(P_1+P_2)^2\,, \qquad
   \tau = \frac{m_\phi^2+q_T^2}{s} \,,
\end{equation}
Generally, the differential cross section can be written as
\begin{equation}\label{eq:standardcrosec}
\begin{aligned}
   d\sigma = \frac{1}{2s} \frac{d^3\vec{q}}{(2\pi)^32E_\phi}\int d^4 x \langle N_1(P_1) N_2(P_2)|{\hat \Phi}^\dagger(x)|\phi(q)\rangle\langle \phi(q)|{\hat \Phi}(0)|N_1(P_1) N_2(P_2)\rangle \,
\end{aligned}
\end{equation}
with
\begin{equation}\label{eq:opratorphi}
 {\hat \Phi}=2\sqrt2 K_{ab}^i \phi_i^\dagger \psi_a^T\big(\lambda_L^*P_R+\lambda_R^*P_L\big)\psi_b\,.
\end{equation}

In SCET, the $n$-collinear quark $\psi_n$ can be written as 
\begin{eqnarray}\label{colfield}
 \chi_n(x) &=& W^\dagger_n (x) \xi_n(x),\quad \xi_n(x)=\frac{\nslash\bnslash}{4}\psi_n(x),
\end{eqnarray}
where $W_n(x)$ is the $n$-collinear Wilson line~\cite{Bauer:2001yt}, which describes the emission of arbitrary $n$-collinear gluons from an $n$-collinear quark.

At the leading power in $\lambda$, only the $n\mcdot A_s$ component of soft gluons can interact with the $n$-collinear field. Such interaction is eikonal and can be removed by a field redefinition~\cite{Bauer:2001yt}:
\begin{eqnarray}
\label{eqs:frd}
 \chi_n(x) \to Y_n(x)\chi_n(x),\qquad
 \phi_v(x) \to Y_v(x)\phi_v (x)\,,
\end{eqnarray}
with
\begin{equation}
 Y_n(x) = \mathbf{P} \exp\left( ig_s\int^0_{-\infty}ds\,n\mcdot A^a_s(x+sn)t^a\right)\,,
\end{equation}
and
\begin{equation}
 Y_v(x) = \mathbf{P} \exp\left(-ig_s\int^{\infty}_0ds\,v\mcdot A^a_s(x+sv)t^a\right)\,,
\end{equation}
where $Y_n(x)$ and $Y_v(x)$ are incoming and outgoing Wilson lines~\cite{Bauer:2001yt,Chay:2004zn,Korchemsky:1991zp}, respectively, and $v$ is the dimensionless vector along the directions of the momentum of the massive scalar with $v^2=1$.
After the fields redefinition, the operator ${\hat \Phi}$ can be written as
\begin{eqnarray}
  {\hat \Phi} &\rightarrow& C_S(-q^2-i\varepsilon,\mu){\hat{\cal O}} \,,
\end{eqnarray}
where
\begin{eqnarray}\label{eq:Phimatch}
  {\hat{\cal O}}= 2\sqrt 2 K_{ab}^i
  Y_{v}^\dagger\phi_v^\dagger\chi_{\bar n}^T Y_{\bar n}C \left(\lambda_L^*P_R+\lambda_R^*P_L\right)Y_n\chi_{n} \,,
\end{eqnarray}
and $C_S(-m_\phi^2-i\varepsilon,\mu)$ is the hard Wilson coefficient. $C$ is the charge conjugation matrix. The matrix element for the process of single colored scalar production can factorize in the form
\begin{eqnarray}\label{eq:MatrixEFactor}
 \langle N_1(P_1) N_2(P_2)|{\hat{\cal O}}^\dagger(x) {\hat{\cal O}}(0)|N_1(P_1) N_2(P_2)\rangle &=&
 \frac{2N_D\lambda^2}{N_c^2}
  \langle N_1(P_1)|{\bar\chi}_{n}(x)\frac{\bnslash}{2}\chi_{n}(0)|N_1(P_1)\rangle
  \qquad\qquad \nn \\
  && \times \langle N_2(P_2)|{\bar\chi}_{\bar n}(x)\frac{\nslash}{2}\chi_{\bar n}(0)|N_2(P_2)\rangle
  \,\,{\cal S}(x,\mu)\,,
\end{eqnarray}
where
\begin{equation}\label{eq:Wsoft}
   {\cal S}(x,\mu)=\frac{1}{N_D}\langle 0|\,{\rm Tr}[
   \overline{\bf T}\big(Y_n^\dagger\,Y_{\bar n}^\dagger\,Y_v\big)(x)\,
   {\bf T}\big( Y_{\bar n}\,Y_n\,Y_v^\dagger\big)(0)]|0\rangle
\end{equation}
is the soft function. The trace is over color indices, and the time-ordering operator ${\bf T}$ is required to ensure the proper ordering of soft gluon fields in the soft Wilson line. $N_D$ is the dimension of the color representation of the scalars. The initial collinear sectors in Eq.~(\ref{eq:MatrixEFactor}) can reduce to the transverse momentum dependent parton distribution functions (TMD PDFs) \cite{Becher:2010tm}
\begin{eqnarray}\label{eq:Bdef}
   {\cal B}_{q/N_1}^{n}(z,x_T^2,\mu)
   &=& \frac{1}{2\pi} \int dt\,e^{-izt\bar n\cdot p}\,\langle N_1(p)|\,
    \bar\chi_n(t\bar n+x_\perp)\,\frac{\rlap/\bar n}{2}\,\chi_n(0)\,|N_1(p)\rangle \,,\nn\\
   {\cal B}_{q/N_2}^{\bn}(z,x_T^2,\mu)
   &=& \frac{1}{2\pi} \int dt\,e^{-iztn\cdot p}\,\langle N_2(p)|\,
    \bar\chi_{\bar n}(tn+x_\perp)\,\frac{\rlap/ n}{2}\,\chi_{\bar n}(0)\,|N_2(p)\rangle \,,
\end{eqnarray}
where $x_T^2\equiv -x_\perp^2>0$. Note that ${\cal B}_{q/N_1}^{n}$ and $ {\cal B}_{q/N_2}^{\bn}$ in Eq.~(\ref{eq:Bdef}) are TMD PDFs for quark. Now the matrix element for the process of a colored scalar production is factorized into two collinear sectors and a soft sector, which do not interact with each other. Thus, the differential cross section can be written as
\begin{equation}\label{eq:Bfactnaive}
\begin{aligned}
   \frac{d^2\sigma}{dq_T^2\,dy}
   &= \frac{2\pi N_D\lambda^2(\mu^2)}{N_c^2 s} {\cal H}(m_\phi^2,\mu^2)
    \frac{1}{4\pi} \int\!d^2x_\perp\,e^{-iq_\perp\cdot x_\perp} \\
   &\quad\times \bigg[ {\cal B}_{q/N_1}^n(\xi_1,x_T^2,\mu)\,
    {\cal B}_{q'/N_2}^{\bn}(\xi_2,x_T^2,\mu){\cal S}(x_T^2,\mu) + (q \leftrightarrow q')\bigg]
    + {\cal O}\bigg( \frac{q_T^2}{M^2} \bigg) \,,
\end{aligned}
\end{equation}
where $y$ is the rapidity of the colored scalar, $\xi_{1,2} = \sqrt{\tau}\,e^{\pm y}$, and
$\mathcal{H}(m_\phi^2,\mu^2)$ is the hard function defined as $\mathcal{H}(m_\phi^2,\mu^2)=\left|C_S(-m_\phi^2-i\varepsilon,\mu^2)\right|^2$. The collinear anomalous terms can be factored out~\cite{Becher:2010tm}, and the product of the two TMD PDFs can be refactorized
\begin{equation}
  \mathcal{B}^{n}_{q/N_1}(z_1,x_\perp,\mu) \, \mathcal{B}^{\bar{n}}_{q'/N_2}(z_2,x_\perp,\mu) = \left( \frac{x_T^2m_\phi^2}{4e^{-2\gamma_E}} \right)^{-F_{q q'}(x_T^2,\mu)} B^{n}_{q/N_1}(z_1,x_\perp,\mu) \, B^{\bar n}_{q'/N_2}(z_2,x_\perp,\mu) \, ,
\end{equation}
where $F_{q q'}$ is the same as the $F_{q \bar q} $ in Ref.~\cite{Becher:2010tm}. The $B_{q/N}$ functions are intrinsically non-perturbative objects. For $x_T \ll 1/\Lambda_{\text{QCD}}$, it can be matched onto the normal PDFs \cite{Becher:2010tm} via
\begin{equation}
  \label{eq:Imatch}
  B_{q/N}(z,x_T^2,\mu)=\sum_{i} \int \frac{d\xi}{\xi} \, I_{q \leftarrow i}(\xi,L_\perp,\mu) \, f_{i/N}(z/\xi,\mu)\,,
\end{equation}
with perturbatively calculable matching coefficient functions $I_{i \leftarrow j}$. Now the differential cross section can be further in a useful form
\begin{eqnarray}\label{eq:fact1}
   \frac{d^2\sigma}{dq_T^2\,dy}
   &=& \frac{2\pi N_D\lambda^2(\mu)}{N_c^2 s} {\cal H}(m_\phi^2,\mu^2) \sum_{i,j=q,q',g}
    \int_{\xi_1}^1\!\frac{dz_1}{z_1} \int_{\xi_2}^1\!\frac{dz_2}{z_2} \nn\\
   &&\mbox{}\times C_{q q'\leftarrow i j}(z_1,z_2,q_T^2,m_\phi^2,\mu)
   f_{i/N_1}(\xi_1/z_1,\mu)\,f_{j/N_2}(\xi_2/z_2,\mu)
\end{eqnarray}
with
\begin{equation}\label{eq:Cdef}
\begin{aligned}
   C_{q q'\leftarrow i j}(z_1,z_2,q_T^2,m_\phi^2,\mu)
   &= \frac{1}{4\pi} \int\!d^2x_\perp\,e^{-iq_\perp\cdot x_\perp}
    \left( \frac{x_T^2 m_\phi^2}{b_0^2} \right)^{-F_{q q'}(L_\perp,a_s)} \\
   &\quad\times I_{q\leftarrow i}^n(z_1,L_\perp,a_s)\,I_{q' \leftarrow j}^{\bn}(z_2,L_\perp,a_s)\,
   {\cal S}(L_\perp,a_s) \,,
\end{aligned}
\end{equation}
where $a_s$, $L_\perp$, $b_0$ are defined as
\begin{equation}\label{eq:LTdefs}
   a_s = \frac{\alpha_s(\mu)}{4\pi}\,, \qquad
   L_\perp = \ln\frac{x_T^2\mu^2}{b_0^2} \,, \qquad
   b_0=2e^{-\gamma_E}\,.
\end{equation}

\section{Resummation}\label{sec:resum}
\subsection{Running of the new physics coupling}\label{subsec:lambdarun}
The new physics coupling $\lambda$ satisfies the renormalization group (RG) equation
\begin{equation}\label{eq:lambdaevo}
  \frac{d\,\ln\lambda}{d\,\ln\mu} = \gamma^\lambda(\alpha_s)\,,
\end{equation}
where the one-loop level $\gamma^\lambda$ is given by
\begin{eqnarray}\label{eq:gammalambda}
 \gamma_0^\lambda=-6C_F\,.
\end{eqnarray}
By solving Eq.~(\ref{eq:lambdaevo}), we can get $\lambda$ running from the scale $\mu_\lambda$ to the factorization scale $\mu$
\begin{equation}\label{eq:runlambda}
  \lambda(\mu^2)=e^{-a_{\gamma^\lambda}(\mu_\lambda^2,\mu^2)}\lambda_0\,,
\end{equation}
where $\lambda_0$ denotes the new physics coupling at the scale $\mu_\lambda$. In this paper, we choose $\mu_\lambda=m_\phi$. $a_{\gamma^\lambda}(\nu^2,\mu^2)$ is defined as
\begin{equation}\label{eq:agammaL}
 a_{\gamma^\lambda}(\nu^2,\mu^2) = -\int_{\alpha_s(\nu^2)}^{\alpha_s(\mu^2)}d\alpha \frac{\gamma^\lambda(\alpha)}{\beta(\alpha)}\,.
\end{equation}
Now, the anomalous dimension of the new physics coupling is only available at the NLO, which means that the resummation for $\lambda$ is at the next-to-leading logarithmic (NLL) order.

\subsection{Hard function}\label{subsec:resumHardF}
In SCET, $C_S(-m_\phi^2,\mu^2)$ (here and below the negative arguments are understood with a $-i\varepsilon$ prescription) can be obtained to order ${\cal O}(\alpha_s)$ from one-loop virtual correction calculation, whose infrared divergences are subtracted in the $\overline{\rm MS}$ scheme~\cite{Zhan:2013sza}
\begin{equation}\label{eq:CsFunc}
   C_S(-m_\phi^2,\mu^2)=1+\frac{\alpha_s(\mu)}{4\pi}
   \left[C_F\left(-L^2+\frac{\pi^2}{6}-2\right)+C_D\left(L-\frac{2}{3}\pi^2-1\right)\right]\,,
\end{equation}
with
\begin{equation}\label{eq:LH}
   L=\ln\frac{-m_\phi^2}{\mu^2}\,.
\end{equation}
$C_S(-m_\phi^2,\mu^2)$ satisfies the RG equation ~\cite{Becher:2009kw}
\begin{equation}\label{eq:Csevol}
   \frac{d}{d\ln\mu}\,C_S(-m_\phi^2,\mu^2)
   = \left[ \Gamma_{\rm cusp}^F(\alpha_s)\,\ln\frac{-m_\phi^2}{\mu^2} + 2\gamma^q(\alpha_s) + \gamma^D(\alpha_s) - \gamma^\lambda(\alpha_s)
    \right] C_S(-m_\phi^2,\mu^2) \,.
\end{equation}
$\Gamma_{\rm cusp}^F(\alpha_s)$ is the cusp anomalous dimension in the fundamental representation. $\gamma^q$ (equal to $\gamma^V/2$ in Ref.~\cite{Becher:2007ty}) is the anomalous dimension of massless quark, and $\gamma^D$ is the one of colored scalar, which is given by~\cite{Becher:2009kw}
\begin{eqnarray}\label{eq:adofdiquark}
\gamma_0^D&=&-2\,C_D\,,\nn\\
\gamma_1^D&=&C_DC_A\left(\frac{2\pi^2}{3}-\frac{98}{9}-4\zeta_3\right)+\frac{40}{9}C_DT_Fn_f\,.
\end{eqnarray}
From now on, the coupling $\alpha_s$ without an explicit argument will always refer to $\alpha_s(\mu)$.

The solution of Eq.~(\ref{eq:Csevol}) is
\begin{equation}\label{eq:resummedH}
\begin{aligned}
 C_S(-m_\phi^2,\mu^2)=\exp\bigg[ 2S(\mu_h^2,\mu^2)- a_\Gamma(\mu_h^2,\mu^2)\ln\frac{-m_\phi^2}{\mu_h^2}
 - a_{\gamma^H}(\mu_h^2,\mu^2) + a_{\gamma^\lambda}(\mu_h^2,\mu^2) \bigg]C_S(-m_\phi^2,\mu_h^2)\,,
\end{aligned}
\end{equation}
where $\gamma^H=2\gamma^q+\gamma^D$, $\mu_h$ is hard matching scale and $S(\nu^2,\mu^2)$ is defined as
\begin{equation}\label{eq:SudakovS}
 S(\nu^2,\mu^2)=-\int_{\alpha_s(\nu^2)}^{\alpha_s(\mu^2)}d\alpha \frac{\Gamma_{\rm cusp}^F(\alpha)}{\beta(\alpha)}
 \int_{\alpha_s(\nu^2)}^{\alpha}\frac{d\alpha'}{\beta(\alpha')}\,.
\end{equation}
$a_{\gamma^H}$ and $a_{\gamma^\lambda}$ have the similar expression as (\ref{eq:agammaL}). Up to NNLL, three-loop $\Gamma_{\rm {cusp}}$ and two-loop normal anomalous dimension are required, and the explicit expressions of them are collected in the Appendix of Ref.~\cite{Becher:2007ty}.
\subsection{Soft function}\label{subsec:soft}
\begin{figure}
\begin{center}
\includegraphics[height=0.242\textwidth]{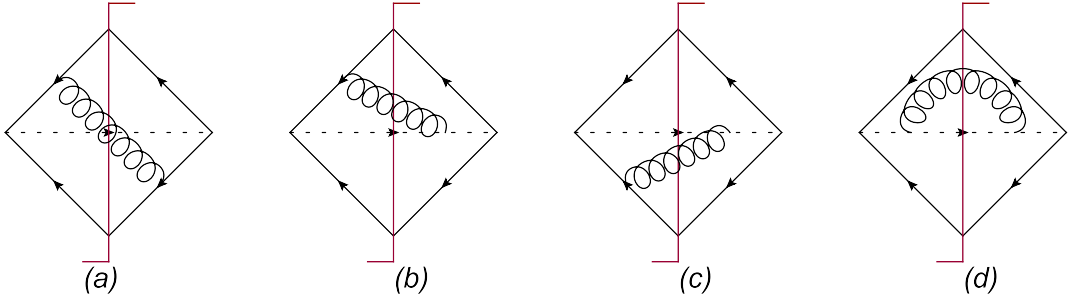}
\vspace{-0.5cm}
\end{center}
\caption{\label{fig:SoftDiagrams}
One-loop diagrams contributing to the soft function ${\cal S}(x_T,\mu)$ .The contributions from diagrams (a), (b), (c) and (d) are denoted as ${\cal S}_a^{(1)}$, ${\cal S}_b^{(1)}$, ${\cal S}_c^{(1)}$ and ${\cal S}_d^{(1)}$. The vertical lines indicate cut propagators.}
\end{figure}
Because the colored scalar in the final state can interact with gluon, the soft function is not trivial any more, which is different from the case of Drell-Yan. At NLO, the diagrams of calculating in eikonal approximation are shown in Fig.~\ref{fig:SoftDiagrams}. In Ref.~\cite{Becher:2010tm}, it has been shown that the contribution from Fig.~\ref{fig:SoftDiagrams}(a) vanishes because the relevant integral is scaleless. The contributions from Fig.~\ref{fig:SoftDiagrams}(b) and Fig.~\ref{fig:SoftDiagrams}(c) are given by
\begin{equation}\label{eq:SCalcbc}
\begin{aligned}
   {\cal S}_b^{(1)}(x_T^2,\mu)
   &= \frac{2\,g_s^2\,\mu^{2\epsilon}}{(2\pi)^{3-2\epsilon}}
     \frac{1}{2}C_D\int d^{4-2\epsilon}k\,\delta(k^2)\theta(k^0)
     e^{-ik_\perp\cdot x_\perp}\left(\frac{\nu_2}{\bn\cdot k}\right)^\beta
     \frac{n\cdot v}{(n\cdot k)(v\cdot k)}\,,\\
   {\cal S}_c^{(1)}(x_T^2,\mu)
   &= \frac{2\,g_s^2\,\mu^{2\epsilon}}{(2\pi)^{3-2\epsilon}}
     \frac{1}{2}C_D\int d^{4-2\epsilon}k\,\delta(k^2)\theta(k^0)
     e^{-ik_\perp\cdot x_\perp}\left(\frac{\nu_2}{\bn\cdot k}\right)^\beta
     \frac{\bn \cdot v}{(\bn \cdot k)(v\cdot k)}\,,
\end{aligned}
\end{equation}
where the analytic regularization method \cite{Becher:2010tm} is used. After calculating the integrals in Eq.~(\ref{eq:SCalcbc}), we find that ${\cal S}_b^{(1)}+{\cal S}_c^{(1)}$ is equal to zero. Therefore the soft function only depends on the contribution from Fig.~\ref{fig:SoftDiagrams}(d)
\begin{equation}\label{eq:SCalcd}
\begin{aligned}
  {\cal S}^{(1)}(x_T^2,\mu)&= {\cal S}_d^{(1)}(x_T^2,\mu)\\
   &= -\frac{g_s^2\,\mu^{2\epsilon}}{(2\pi)^{3-2\epsilon}}
     C_D\int d^{4-2\epsilon}k\,\delta(k^2)\theta(k^0)
     e^{-ik_\perp\cdot x_\perp}\left(\frac{\nu_2}{\bn\cdot k}\right)^\beta
     \frac{v^2}{(v\cdot k)^2}\,,
\end{aligned}
\end{equation}
and in the $\overline{\rm {MS}}$ scheme, the NLO soft function is
\begin{equation}\label{eq:Sresult}
\begin{aligned}
   {\cal S}(x_T^2,\mu) = 1+\frac{\alpha_s C_D}{2\pi}\ln \frac{x_T^2\mu^2}{4e^{-2\gamma_E}}+{\cal O}(\alpha_s^2)\,.
\end{aligned}
\end{equation}
The RG equation of the soft function is
\begin{equation}\label{eq:Sevol}
\begin{aligned}
   \frac{d\ln {\cal S}(x_T^2,\mu)}{d\ln\mu}= 2\gamma^{S_{qq}}(\alpha_s) \,.
\end{aligned}
\end{equation}
where $\gamma^{S_{qq}}$ is the anomalous dimension of the soft function, which can be obtained at one-loop level from Eq.~(\ref{eq:Sresult}).

In Ref.~\cite{Becher:2011xn},  the double logarithmic terms of the function $I_{q\leftarrow q}$ can be resummed by defining a new function $\bar I_{q\leftarrow q}$. The logarithmic term in the soft function can be resummed in the same way,
\begin{equation}\label{eq:Sredef}
   {\cal S}(L_\perp,a_s)\equiv e^{h_S(L_\perp,a_s)}\bar{\cal S}(L_\perp,a_s) \,.
\end{equation}
From the RG equation of the soft function (\ref{eq:Sevol}), we can obtain
\begin{eqnarray}\label{eq:hsevol}
   \frac{d}{d\ln\mu}\,h_S(L_\perp,a_s(\mu))&=& 2\gamma^{S_{qq}}(a_s(\mu)) \,,\nn\\
   \frac{d}{d\ln\mu}\,{\bar{\cal S}}(L_\perp,a_s(\mu))&=&0
\end{eqnarray}
Matching to the NLO result of ${\cal S}(L_\perp,a_s(\mu))$, we can get ${\bar{\cal S}}(L_\perp,a_s(\mu))={\cal S}(0,a_s(\mu))$ by choosing the boundary condition as $h_S(0,a_s(\mu))\equiv 0$. Generalizing the NLO result to high orders, $h_S(L_\perp,a_s(\mu))$ can be expanded
\begin{equation}\label{eq:hssum}
   h_S(L_\perp,a_s(\mu)) = \sum_{n=1}^\infty\,h_S^{(n)}(L_\perp)a_s(\mu)^n \,.
\end{equation}
Using the RG equation of the soft function, we get the first two expansion coefficients of $h_S$
\begin{eqnarray}\label{eq:2hscoeff}
  h_S^{(1)}(L_\perp)&=&\gamma_0^{S_{qq}}\,L_\perp\,,\nn \\
  h_S^{(2)}(L_\perp)&=&\frac{1}{2}\gamma_0^{S_{qq}}\,\beta_0\,L_\perp^2
   +\gamma_1^{S_{qq}}\,L_\perp\,.
\end{eqnarray}
\subsection{Scale independence}
In the factorization formalism, we have introduced the hard and soft function. It is important to check the scale independence of the final results at one-loop level. As shown in Ref.~\cite{Becher:2010tm}, the RG equation for the PDFs is
\begin{equation}
   \frac{d}{d\ln\mu}\,f_{i/N}(z,\mu)
   = \sum_j \int_z^1\!\frac{du}{u}\,{\cal P}_{i\leftarrow j}(z/u,\mu)\,f_{j/N}(u,\mu) \,,
\end{equation}
the evolution equations for the kernel function $I_{q\leftarrow i}(z,x_T^2,\mu)$ are
\begin{equation}\label{eq:Ievol}
\begin{aligned}
   \frac{d}{d\ln\mu}\,I_{q\leftarrow i}(z,x_T^2,\mu)
   &= \left[ \Gamma_{\rm cusp}^F(\alpha_s)\,\ln\frac{x_T^2\mu^2}{4 e^{-2\gamma_E}}
    - 2\gamma^q(\alpha_s) \right] I_{q\leftarrow i}(z,x_T^2,\mu) \\
   &\quad\mbox{}- \sum_j \int_z^1\!\frac{du}{u}\,I_{q\leftarrow j}(u,x_T^2,\mu)\,
    {\cal P}_{j\leftarrow i}(z/u,\mu) \,,
\end{aligned}
\end{equation}
and the RG equation for $F_{qq'}(x_T^2,\mu)$ is
\begin{equation}
\begin{aligned}
\frac{dF_{qq'}(x_T^2,\mu)}{d\ln\mu}= 2\Gamma_{\rm cusp}^F(\alpha_s)\,.
\end{aligned}
\end{equation}
The RG invariance requires
\begin{equation}
\begin{aligned}
\frac{d}{d\ln\mu}\left[\lambda^2(\mu) {\cal H}(\mu^2)
\left(\frac{x_T^2 m_\phi^2}{b_0^2} \right)^{-F_{q q'}(L_\perp)}
I_{q\leftarrow i}^n(L_\perp)\,I_{q' \leftarrow j}^{\bn}(L_\perp) f_{i/N_1}(\mu)\,f_{j/N_2}(\mu)\,{\cal S}(L_\perp)\right]=0\,,
\end{aligned}
\end{equation}
which implies
\begin{equation}\label{eq:RGinvAD}
\begin{aligned}
\gamma^H-2\gamma^q+\gamma^{S_{qq}}=0\,,
\end{aligned}
\end{equation}
where the RG equations of the new physics coupling $\lambda$ (\ref{eq:lambdaevo}), the hard (\ref{eq:Csevol}) and soft function (\ref{eq:Sevol}) have been used. We confirm that Eq.~(\ref{eq:RGinvAD}) is satisfied at one-loop level, through calculating the anomalous dimension $\gamma^H$ and $\gamma^{S_{qq}}$ up to ${\cal O}(\alpha_s)$.

Using Eq.~(\ref{eq:RGinvAD}), the two-loop anomalous dimension of the soft function can be derived from
\begin{equation}
\begin{aligned}
\gamma^{S_{qq}}=-\gamma^D\,,
\end{aligned}
\end{equation}
where $\gamma^D$ is available up to ${\cal O}(\alpha_s^2)$ (\ref{eq:adofdiquark}).
\subsection{Final RG improved differential cross section}
Now, we can obtain the differential cross section of the transverse momentum resummation
\begin{equation}\label{eq:resumx}
\begin{aligned}
   \frac{d^2\sigma}{dq_T^2\,dy}
   =& \frac{2\pi N_D\lambda_0^2}{N_c^2 s} {\cal H}(m_\phi^2,\mu_h^2) U(m_\phi^2,\mu_h^2,\mu^2)\sum_{i,j=q,q',g}
    \int_{\xi_1}^1\!\frac{dz_1}{z_1} \int_{\xi_2}^1\!\frac{dz_2}{z_2} \nn\\
   &\mbox{}\times C_{q q'\leftarrow i j}(z_1,z_2,q_T^2,m_\phi^2,\mu)
   f_{i/N_1}(\xi_1/z_1,\mu)\,f_{j/N_2}(\xi_2/z_2,\mu) \,,
\end{aligned}
\end{equation}
with
\begin{equation}\label{eq:Udef}
U(m_\phi^2,\mu_h^2,\mu^2)=\exp\Bigg[4S(\mu_h^2,\mu^2)- 2a_\Gamma(\mu_h^2,\mu^2)\ln\frac{m_\phi^2}{\mu_h^2}
 - 2a_{\gamma^H}(\mu_h^2,\mu^2) - 2a_{\gamma^\lambda}(\mu_\lambda^2,\mu_h^2)\Bigg]\,,
\end{equation}
and
\begin{equation}\label{eq:redefC}
\begin{aligned}
   C_{q q'\leftarrow i j}(z_1,z_2,q_T^2,m_\phi^2,\mu)
   &= \frac{1}{2} \int_0^\infty\!dx_T\,x_T\,J_0(x_T q_T)\,
    \exp\Big[ g_F(m_\phi^2,\mu,L_\perp,a_s) + h_S(L_\perp,a_s)\Big] \\
   &\quad\times \bar I_{q\leftarrow i}^n(z_1,L_\perp,a_s)\,
    \bar I_{q\leftarrow j}^{\bn}(z_2,L_\perp,a_s) {\bar{\cal S}}(L_\perp,a_s)\,,
\end{aligned}
\end{equation}
where $J_0$ is the zeroth order Bessel function. The expressions of $ \bar I_{q\leftarrow i}$ and $g_F$ have been shown in Ref.~\cite{Becher:2011xn}.

Table \ref{tab:adcounting} shows the counting scheme for resummation \cite{Becher:2007ty}. Up to NNLL, all the required anomalous dimensions are available, except for the two-loop $\gamma^\lambda$. It can only be obtained from the calculation of two-loop $\beta$ function of the new physics coupling $\lambda$, the result of which is not available and need to be studied in the future. Thus, we just use the one-loop $\gamma^\lambda$ in this paper. Actually, the contribution from $\gamma^\lambda$ to the evolution function $U(m_\phi^2,\mu_h^2,\mu^2)$ vanishes when $\mu_h^2\sim m_\phi^2$, so $\gamma^\lambda$ only affects the running of $\lambda(\mu_\lambda^2)$, and our resummation is called as ${\rm NNLL}_{\rm approx}$.
\begin{table}
\begin{center}
\begin{tabular}{ccccc}
\hline
 \ \ Log.\ approx.\ \ &\ \ Accuracy $\sim\alpha_s^n L^k$ \ \
 &\ \ $\Gamma_{\rm cusp} \ \ $
 &\ \ $\gamma^D$, $\gamma^q$, $\gamma^\lambda$ \ \ &\ \ $C_S$, ${\cal S} \ \ $\\
\hline
 LL & $k= 2n$ & 1-loop
 & tree-level & tree-level \\
 NLL & $2n-1\le k\le 2n$& 2-loop
 & 1-loop & tree-level \\
 NNLL & $2n-3\le k\le 2n$ & 3-loop & 2-loop
 & 1-loop \\
\hline
\end{tabular}
\end{center}
\caption{Schemes for resummation with different level of accuracy.
\label{tab:adcounting}
}
\end{table}

To give precise prediction, we resum the singular terms to all orders and include the non-singular terms up to the NLO, which can be written as
\begin{equation}
   \frac{d\sigma^{\rm NNLL_{approx}+NLO}}{dq_T}
   = \frac{d\sigma^{\rm NNLL_{approx}}}{dq_T} + \left( \frac{d\sigma^{\rm NLO}}{dq_T}
    - \frac{d\sigma^{\rm NNLL_{approx}}}{dq_T} \bigg|_{\text{expanded to NLO}} \right) \,.
\end{equation}

\section{The $q_T$ spectrum of colored scalar at fixed order}\label{sec:qtfo}
To verify the correctness of our factorization formula and soft functions, we expand our $q_T$ spectrum to the NLO and compare with the exact NLO results. By expanding $C_{q q'\leftarrow i j}$ to order ${\cal O}(\alpha_s)$ in the limit $q_T\rightarrow 0$, the differential cross section can be written as
\begin{equation}\label{eq:SCETexpand}
\begin{aligned}
 \frac{d^2\sigma}{dq_T^2\,dy}=&\frac{2\pi N_D\lambda^2}{N_c^2 s}
  \Bigg\{f_{q/N_1}(\xi_1)f_{q'/N_2}(\xi_2)
  \left(A\left[\frac{1}{q_T^2}\ln\frac{m_\phi^2}{q_T^2}\right]_\star^{[q_T^2,\mu^2]}
  +B\left[\frac{1}{q_T^2}\right]_\star^{[q_T^2,\mu^2]}+C\delta(q_T^2)\right)\\
  &+\left[\sum_{a}\left(\frac{\alpha_s}{4\pi}\right)
  \left(\frac{1}{2}P_{q\leftarrow a}\left[\frac{1}{q_T^2}\right]_\star^{[q_T^2,\mu^2]}
  +R_{q\leftarrow a}\delta(q_T^2)\right)\otimes f_{a/N_1}\right](\xi_1)f_{q'/N_2}(\xi_2)\\
  &+f_{q/N_1}(\xi_1)\left[\sum_{a}\left(\frac{\alpha_s}{4\pi}\right)
  \left(\frac{1}{2}P_{q'\leftarrow a}\left[\frac{1}{q_T^2}\right]_\star^{[q_T^2,\mu^2]}
  +R_{q'\leftarrow a}\delta(q_T^2)\right)\otimes f_{a/N_2}\right](\xi_2)\\
  &+ (q\leftrightarrow q') \Bigg\}\,,
\end{aligned}
\end{equation}
with
\begin{equation}
\begin{aligned}
A=\frac{\alpha_s}{4\pi}4C_F,\quad
B=-\frac{\alpha_s}{4\pi}(6C_F+2C_D),\quad
C={\cal H}^{(1)}{\cal S}^{(0)}+{\cal H}^{(0)}{\cal S}^{(1)}|_{L_\perp\to 0}\,,
\end{aligned}
\end{equation}
where $P_{q\leftarrow a}$ are the NLO DGLAP splitting functions:
\begin{equation}\label{eq:splittingF}
P_{q\leftarrow q}(z)=4C_F\left(\frac{1+z^2}{1-z}\right)_+\,,\qquad
P_{q\leftarrow g}(z)=4T_F [z^2+(1-z)^2]\,,
\end{equation}
and the remainder functions $R_{q\leftarrow a}$ are
\begin{equation}\label{eq:remainderfuncs}
R_{q\leftarrow q}(z)=C_F\left(2(1-z)-\frac{\pi^2}{6}\delta(1-z)\right)\,,\qquad
R_{q\leftarrow g}(z)=4T_F\,z(1-z)\,.
\end{equation}
The star distribution in Eq.~(\ref{eq:SCETexpand}) is defined as \cite{Bosch:2004th,DeFazio:1999sv}
\begin{eqnarray}\label{eq:starfdef}
\begin{aligned}
\left[f(x)\right]_\star^{[x,a]}&=f(x)\quad  {\rm for}\quad x>0\,,\\
\int_0^a dx\,\left[f(x)\right]_\star^{[x,a]}g(x)&=\int_0^a dx\,f(x)\left[g(x)-g(0)\right]\,.
\end{aligned}
\end{eqnarray}

\begin{figure}[t]
\begin{center}
\begin{tabular}{cc}
\includegraphics[width=0.50\textwidth]{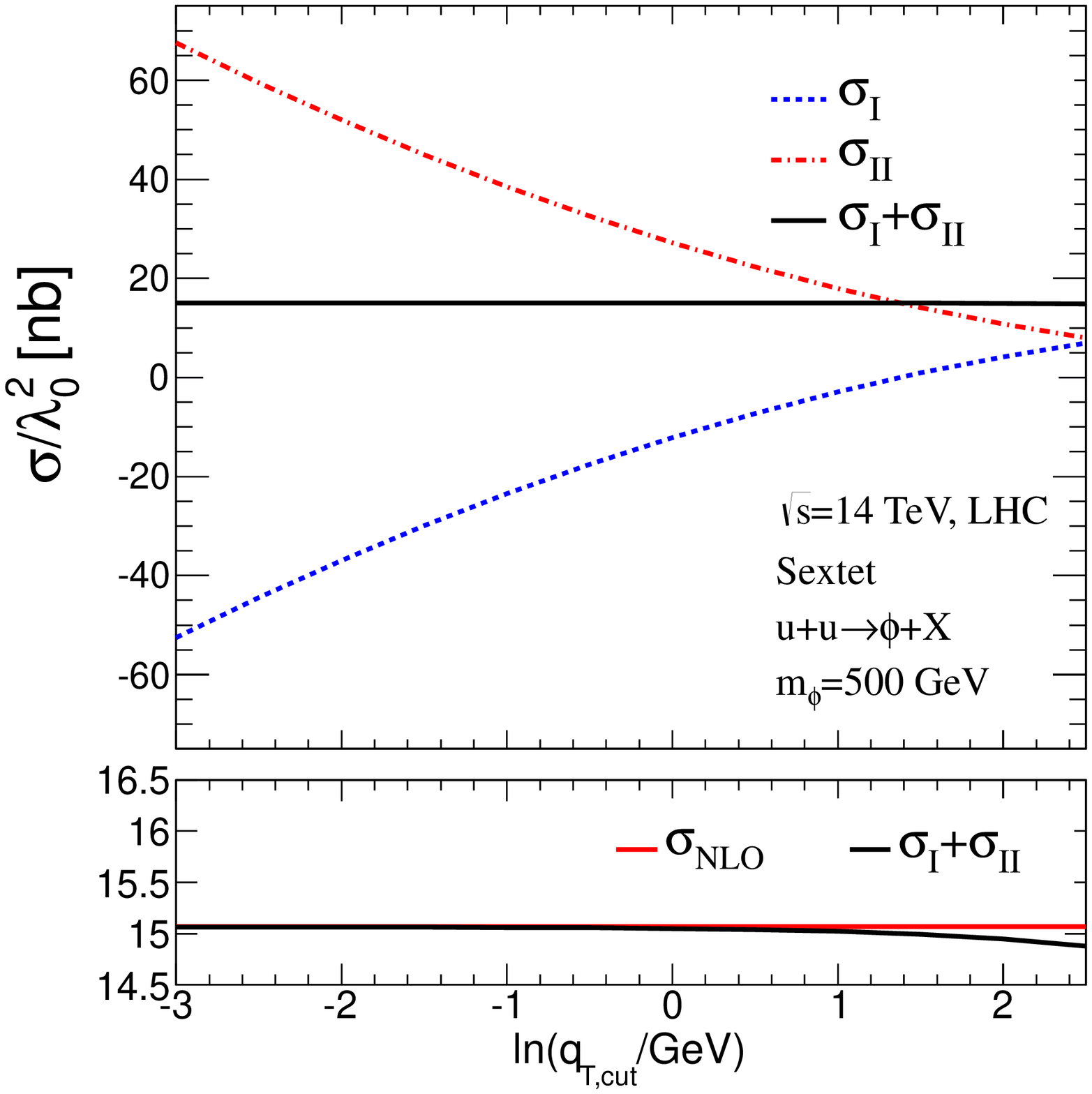} &
\includegraphics[width=0.50\textwidth]{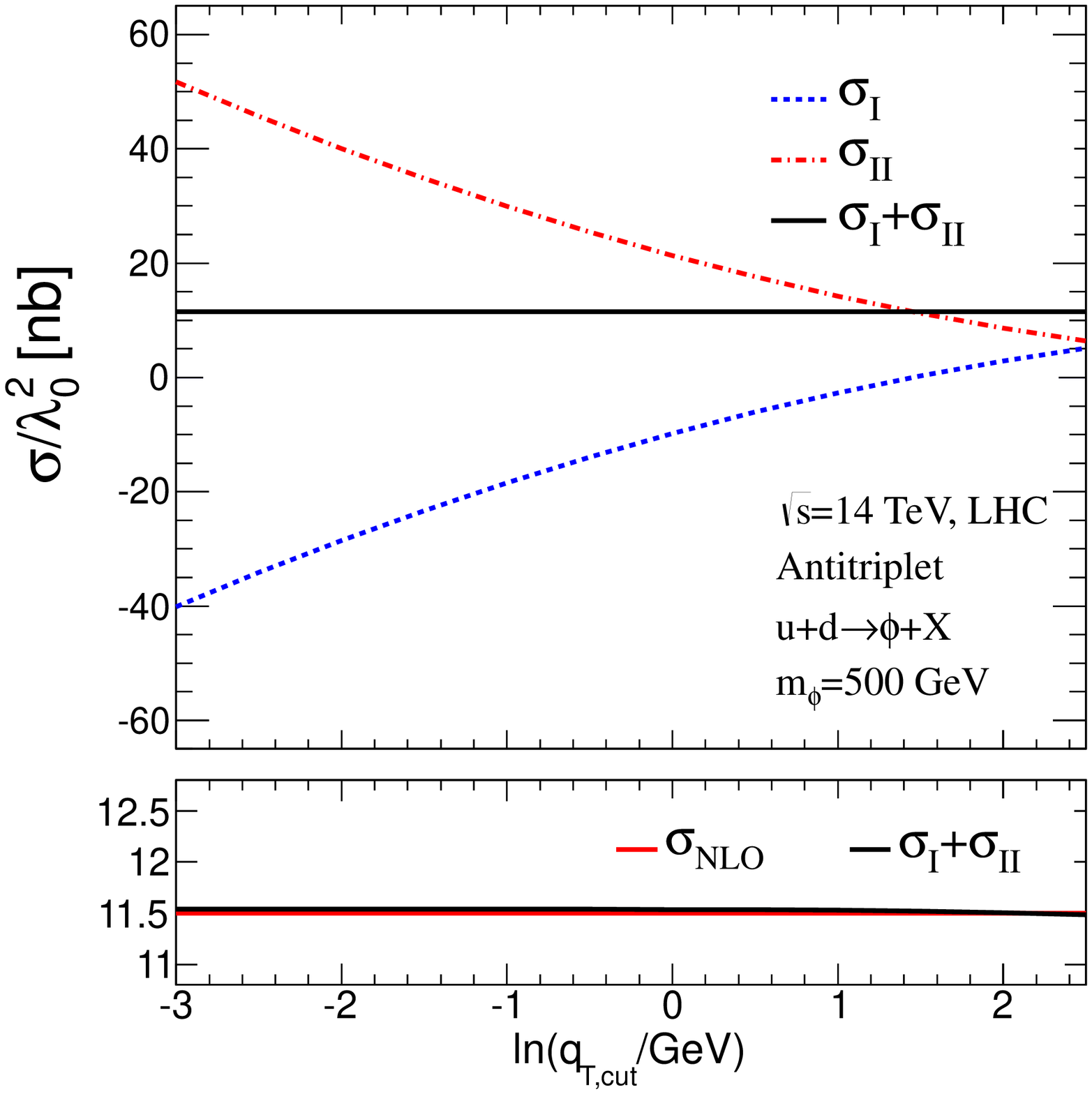}
\end{tabular}
\end{center}
\vspace{-0.5cm}
\caption{\label{fig:compareNLO}
The NLO total cross section for colored scalar production at the LHC with $\sqrt{s}=14{\rm TeV}$. CTEQ6.1 \cite{Stump:2003yu} PDF sets are used. In the lower plots, the red solid lines represent the results in Ref.~\cite{Han:2009ya}.}
\end{figure}

Now we try to reproduce the NLO total cross section for colored scalar production. Using the phase space slicing method, the NLO total cross section can be divided into two parts: small $q_T$ region denoted by $\sigma_{\rm I}$, which can be obtained by integrating the differential cross section in Eq.~(\ref{eq:SCETexpand}) in the approximation of neglecting ${\cal O}(q_T^2/m_\phi^2)$ terms, and the large $q_T$ part denoted by $\sigma_{\rm II}$, which is infrared safe and can be numerically computed directly. Thus the total cross section is given by
\begin{equation}\label{eq:qTcutX}
\begin{aligned}
\sigma_{\rm NLO}=
\int_0^{q_{T,{\rm cut}}^2}dq_T^2\frac{d\sigma_{\rm NLO}}{dq_T^2}
+\int_{q_{T,{\rm cut}}^2}^\infty dq_T^2\frac{d\sigma_{\rm NLO}}{dq_T^2}
=\sigma_{\rm I}+\sigma_{\rm II}\,.
\end{aligned}
\end{equation}
As shown Fig.~\ref{fig:compareNLO}, our numerical results indicate the correctness of the hard and soft function. It can be seen that the dependence on $q_{T,{\rm cut}}$ is canceled after summing $\sigma_{\rm I}$ and $\sigma_{\rm II}$, and the NLO total cross section is agreement with the one in Ref.~\cite{Han:2009ya}.

\section{Numerial discussion}\label{sec:numdis}
In this section, we give the numerical results for the transverse momentum resummation effects in the single production of the color sextet (antitriplet) scalars at the LHC. Throughout the numerical calculation, we use MSTW2008NLO \cite{Martin:2009iq} PDF sets for $\rm NLO$ and $\rm NLL$, and use MSTW2008NNLO PDF sets for $\rm NNLL_{approx}$. In addition, we factored out the new physics coupling $\lambda_0^2$ for a model independent presentation and choose the initial state quarks $uu$ for sextet and $ud$ for antitriplet, respectively. We choose the factorization scale~\cite{Becher:2011xn}
\begin{equation}\label{eq:muchoice}
   \mu=q_T+q_* \,,
\end{equation}
where
\begin{equation}\label{eq:qstar}
   q_* = m_\phi\exp\left(-\frac{2\pi}{\Gamma_0^F\alpha_s(q_*)}\right)\,.
\end{equation}
From the Eq.~(\ref{eq:qstar}), we obtain $q_*= 2.9\,$GeV for $m_\phi=500\,$GeV and $q_*=3.8\,$GeV for $m_\phi=1\,$TeV, both of which are short-distance scales in the perturbative domain.

\begin{figure}[t]
\begin{center}
\begin{tabular}{cc}
\includegraphics[width=0.49\textwidth]{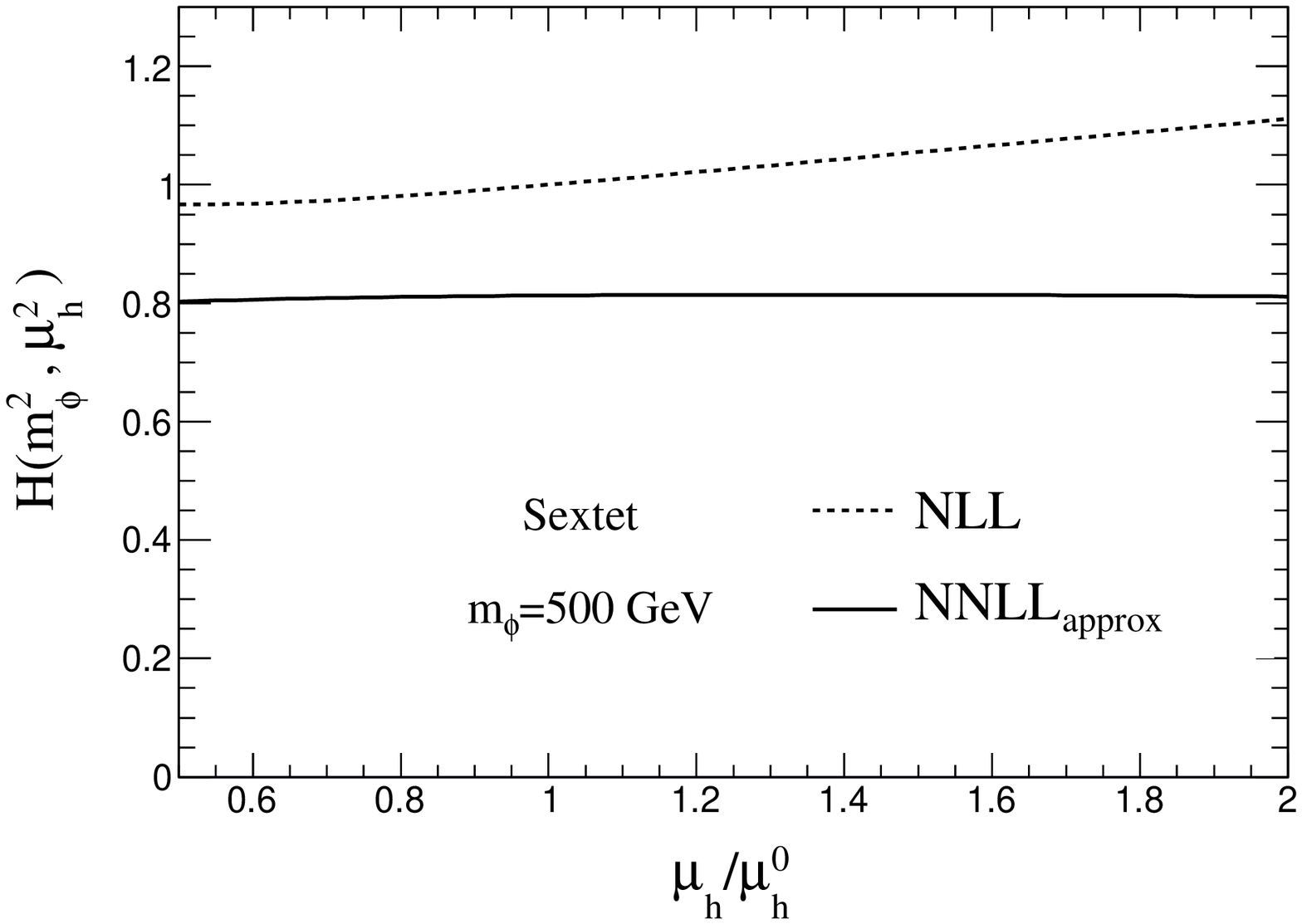} &
\includegraphics[width=0.49\textwidth]{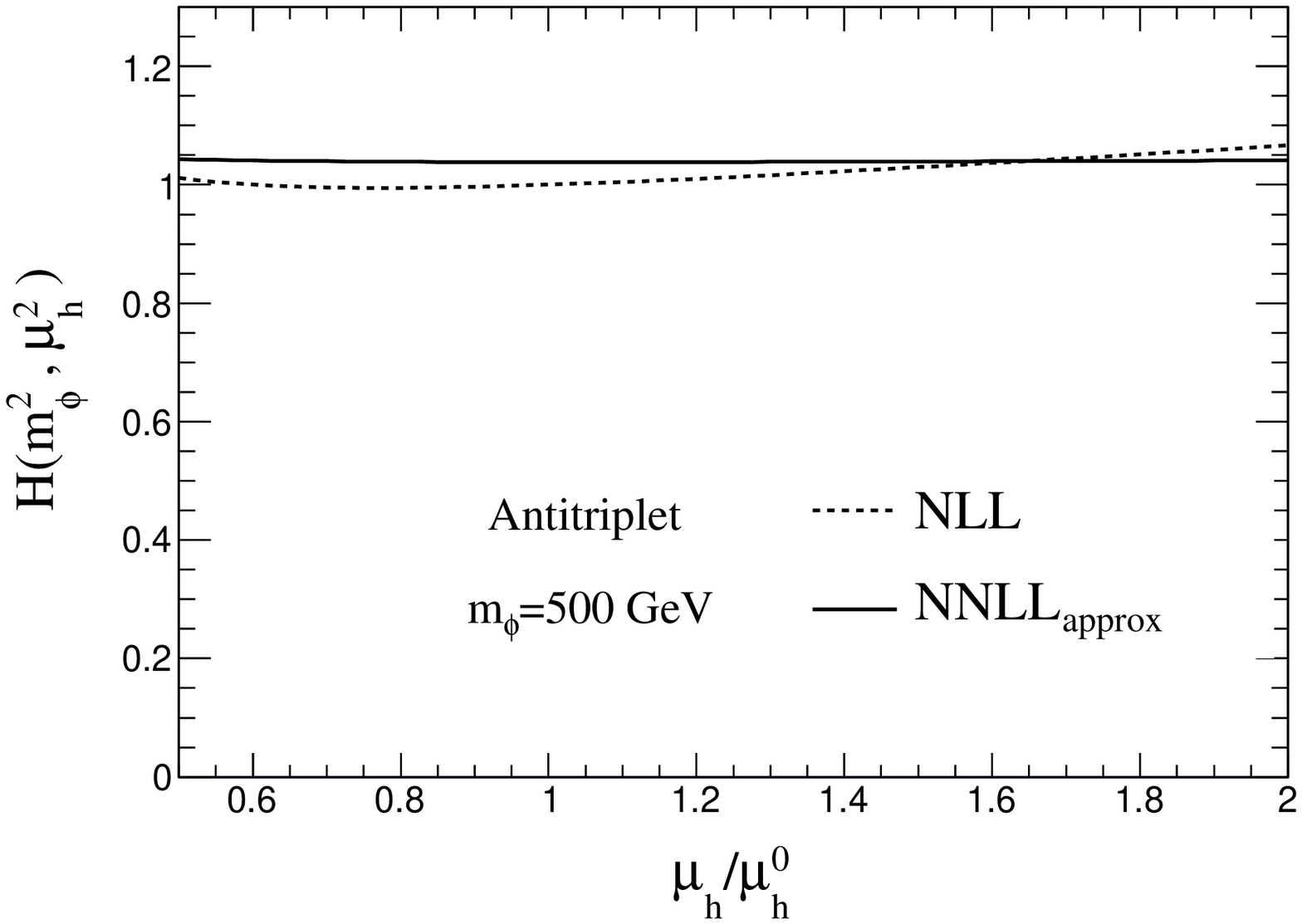}
\end{tabular}
\end{center}
\vspace{-0.5cm}
\caption{\label{fig:muhdep}
Dependence of the hard function on hard matching scale $\mu_h$. For comparison, we show the dependence on $\mu_h$ for both NLL (dashed) and $\rm NNLL_{approx}$ (solid) resummation.}
\end{figure}

Besides, we choose the hard matching scale $\mu_h^0=m_\phi$ for both color sextet and antitriplet. Fig.~\ref{fig:muhdep} shows the dependence of the hard function on $\mu_h$. It can be seen that the hard matching scale dependence decreases significantly from NLL to $\rm NNLL_{approx}$ for both color sextet and antitriplet.

\begin{figure}[htb]
\begin{center}
\begin{tabular}{cc}
\includegraphics[width=0.42\textwidth]{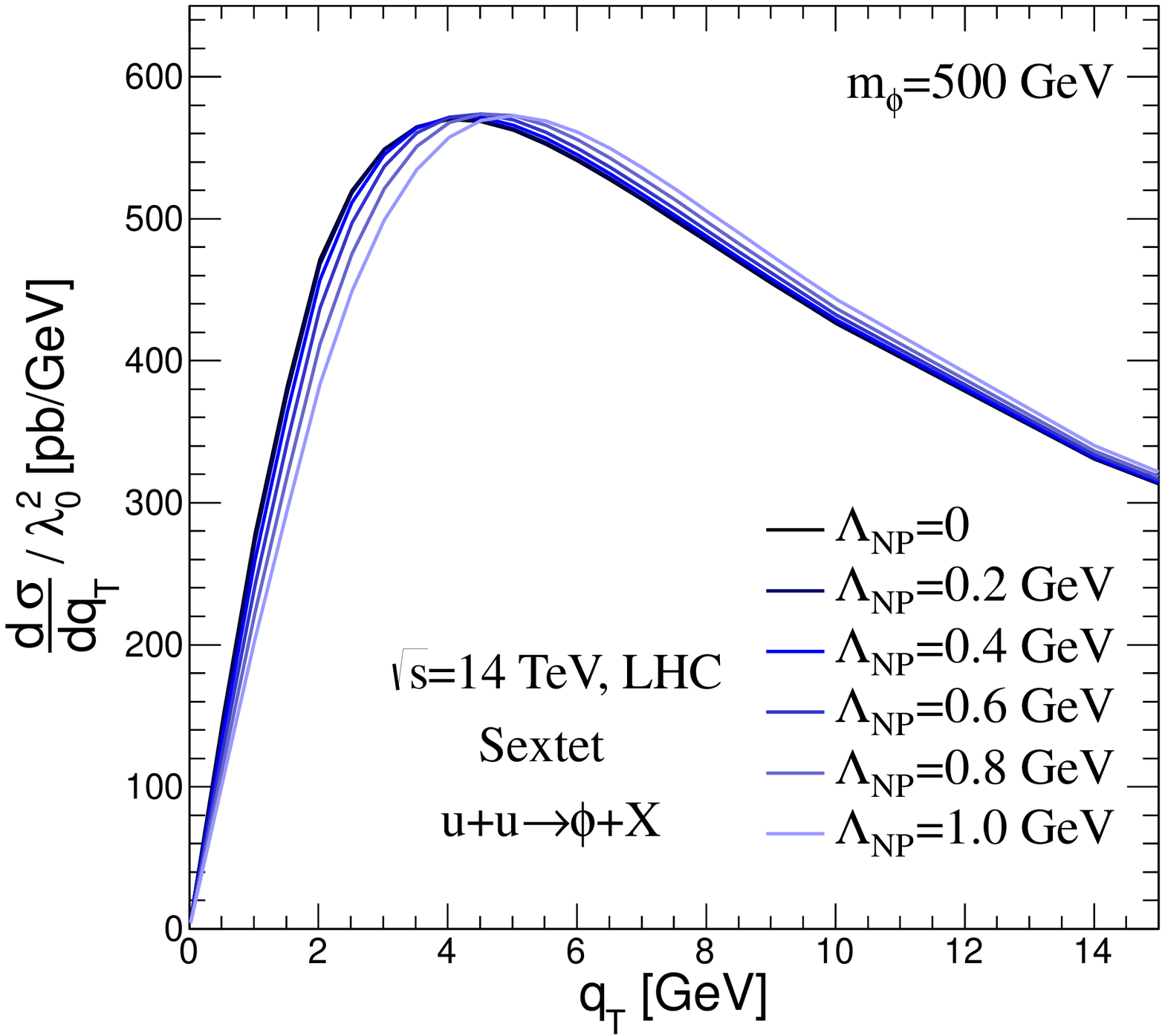}  &
\includegraphics[width=0.42\textwidth]{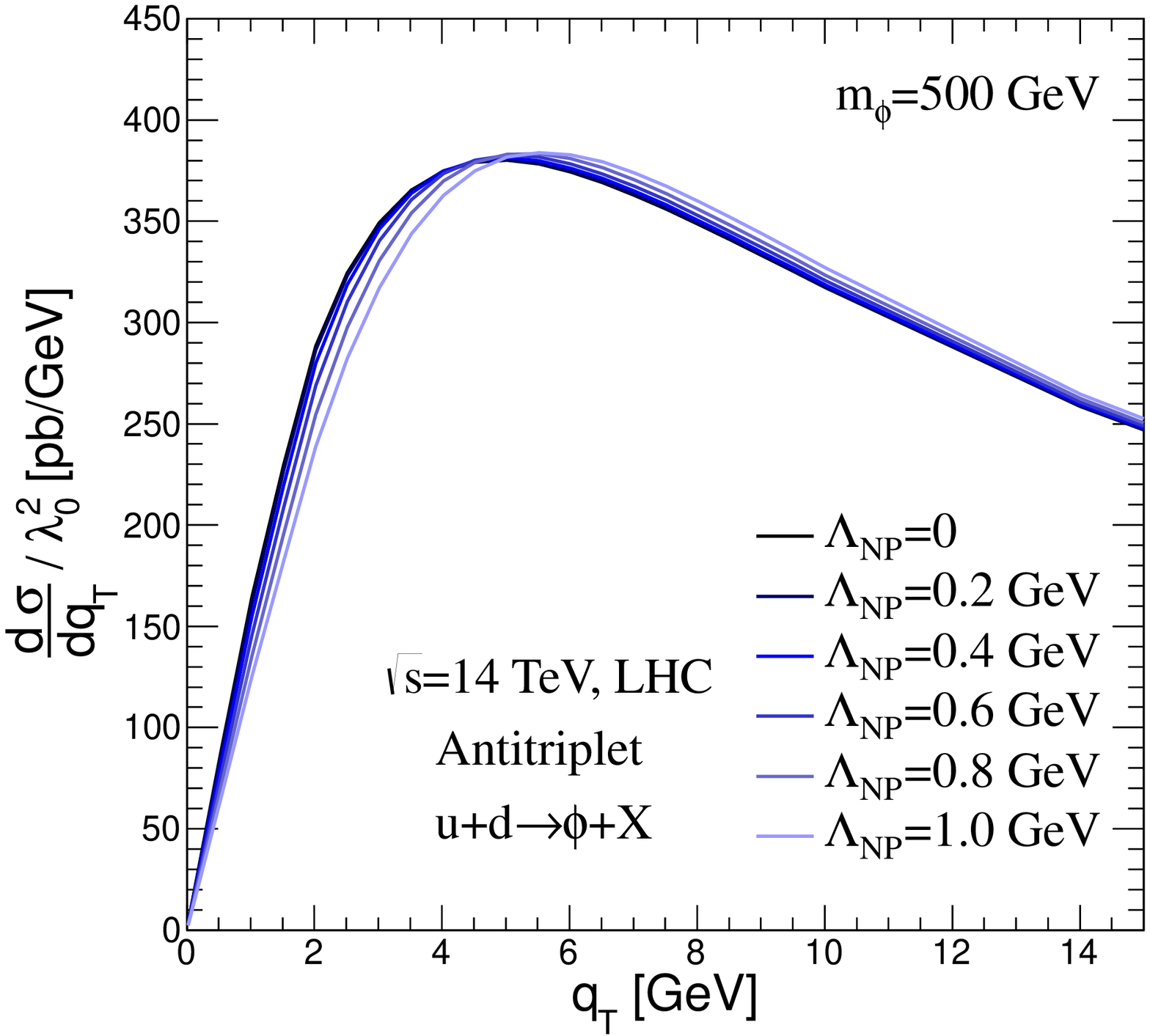}

\end{tabular}
\vspace{-0.5cm}
\end{center}
\caption{\label{fig:NPdis}
Long-distance effects on the differential cross sections $d\sigma /dq_T$ ($\rm NNLL_{approx}$). The left figure is for sextet and the right one is for antitriplet.}
\end{figure}

Our factorization formula is formally valid in the region $\Lambda_{\text{QCD}} \ll q_T \ll m_\phi$. When $q_T \sim \Lambda_{\rm QCD}$,  there are corrections in powers of $x_T\Lambda_{\rm QCD}$, which  comes form the operator-product expansion of the transverse PDFs \cite{Becher:2011xn}. These power corrections are of non-perturbative origin and one must model them, using some technique shown as follows ~\cite{Becher:2011xn}. The TMD PDFs are replaced by
\begin{equation}\label{eq:BiNP}
   B_{q/N}(\xi,x_T^2,\mu)
   = f_{\rm hadr}(x_T\Lambda_{\rm NP})\,B_{q/N}^{\rm pert}(\xi,x_T^2,\mu)\, \,,
\end{equation}
where $\Lambda_{\rm NP}$ is a hadronic scale, and $f_{\rm hadr}(x_T \Lambda_{\rm NP})$ is
\begin{equation}
  f_{\rm hadr}(x_T \Lambda_{\rm NP}) = \exp(-\Lambda_{\rm NP}^2 x_T^2) \, .
\end{equation}

Fig.~\ref{fig:NPdis} shows the $\Lambda_{\rm NP}$ dependence of the results of transverse momentum resummation of single colored scalar production at the LHC with $\sqrt s=14\,\rm TeV$. The non-perturbative form factor results in a small shift of the position of the peak of the $q_T$ distribution. In addition, the $q_{\rm peak}$ of color antitriplet is a little larger than the one of sextet. In the following calculation, we choose $\Lambda_{\rm NP}=600\,{\rm MeV}$ \cite{Becher:2011xn} to simulate the non-perturbative effects for single colored scalar production.

\begin{figure}[htb]
\begin{center}
\begin{tabular}{cc}
\includegraphics[width=0.49\textwidth]{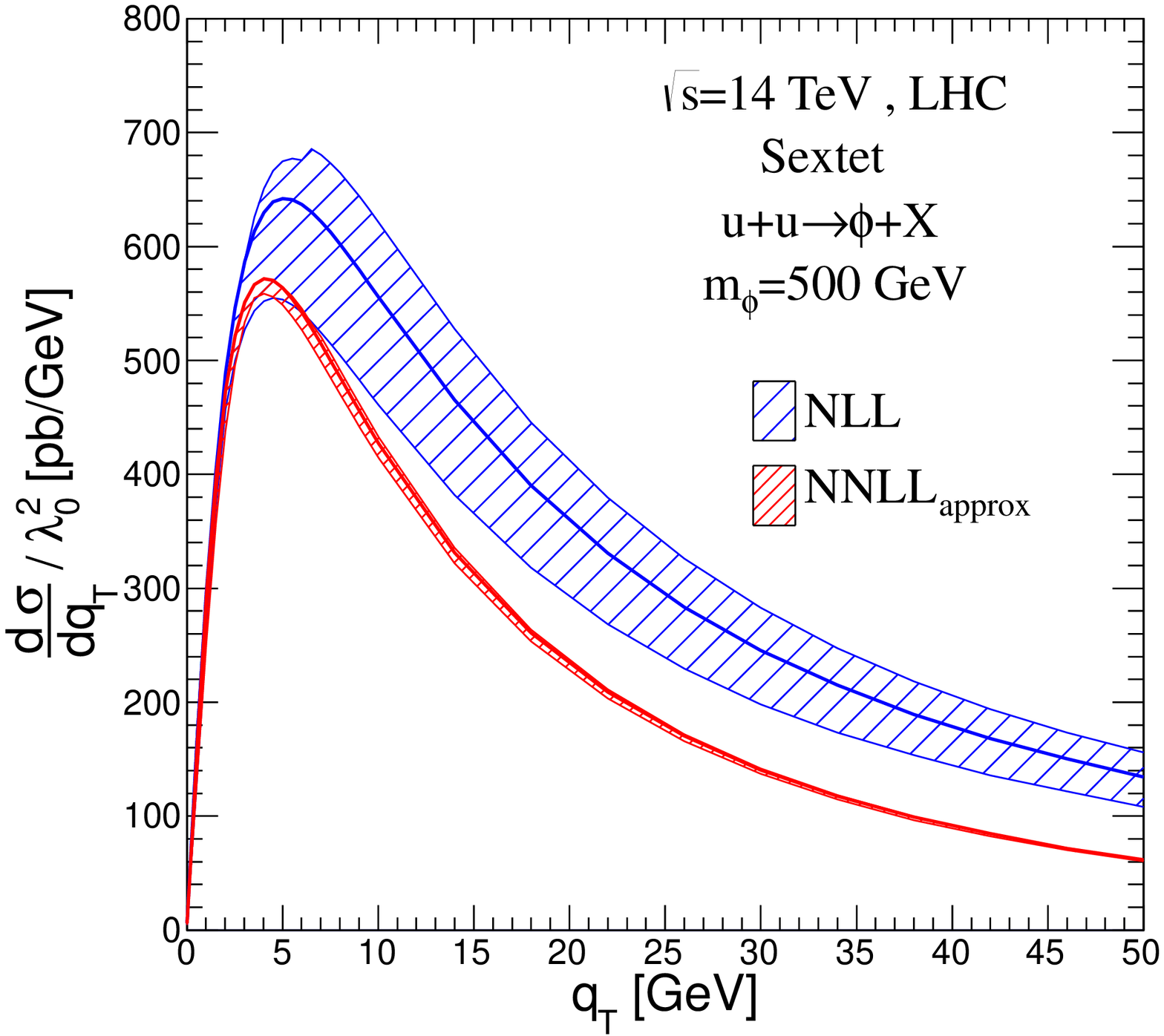} &
\includegraphics[width=0.49\textwidth]{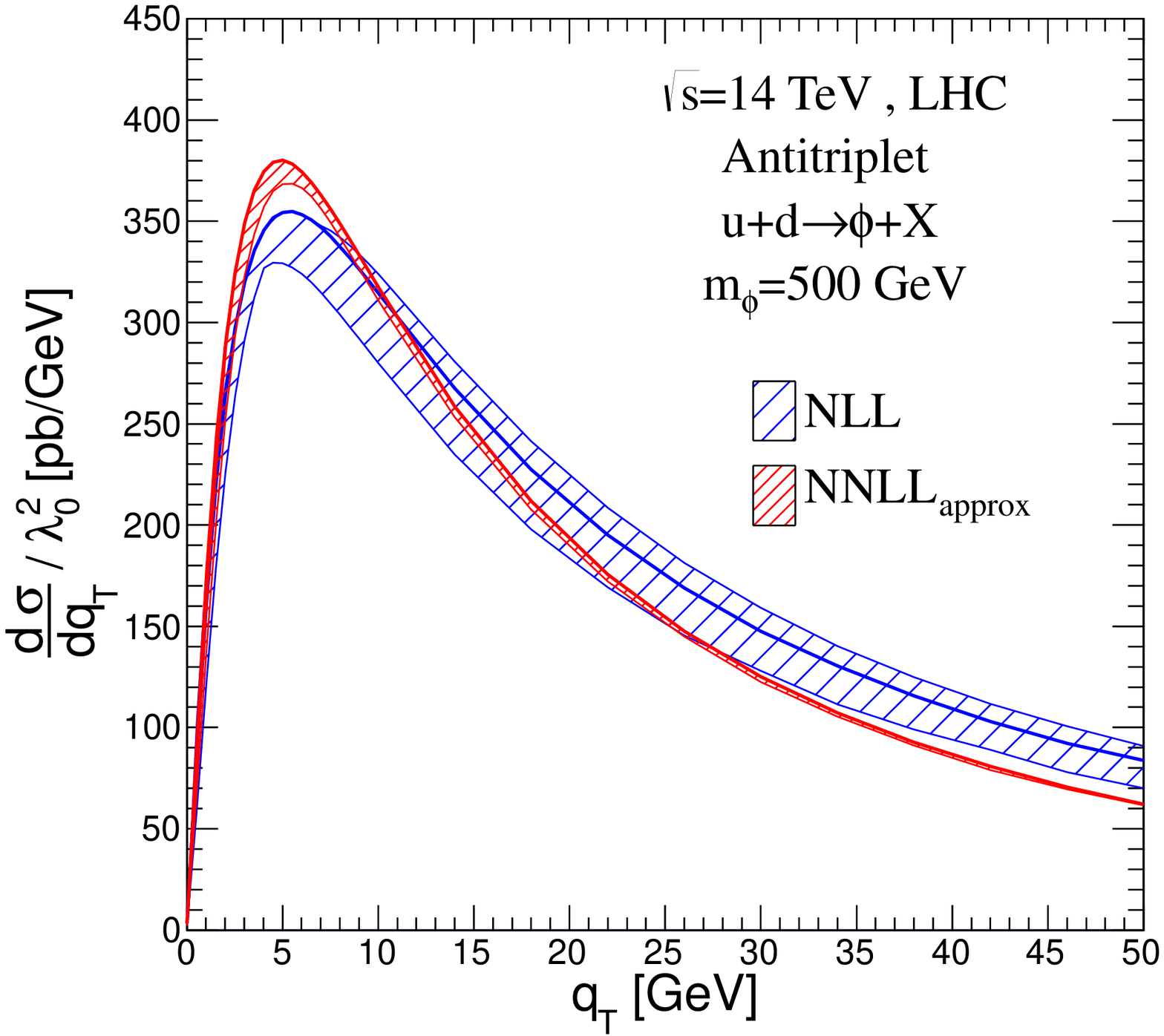}
\end{tabular}
\end{center}
\vspace{-0.5cm}
\caption{\label{fig:scaledep}
 Results of resummation for the transverse momentum distribution of single colored scalar production at $m_\phi=500\,\rm{GeV}$ at the LHC with NLL (blue bands) and $\rm NNLL_{approx}$ (red bands) accuracy. The thick lines represent the default scale choice.
 }
\end{figure}
In Fig.~\ref{fig:scaledep}, we show the scale dependence of the differential cross section at the NLL and the $\rm NNLL_{approx}$, varying the factorization scale $\mu$ by a factor of 2. It can be seen that the scale uncertainties reduce significantly from NLL to $\rm NNLL_{approx}$ for both color sextet and antitriplet. In addition, comparing to NLL, the differential cross section at $\rm NNLL_{approx}$ in the peak region is suppressed for color sextet and enhanced for antitriplet. However, comparing to the color antitriplet case, the $\rm NLL$ result has a larger deviation from the $\rm NNLL_{approx}$ for color sextet. It is because that in $\rm NLL$ calculations the LO hard function is used, while in $\rm NNLL_{approx}$ calculations the NLO hard function is used, which gives a larger NLO correction to LO hard function (about -20\%, as shown in Fig.~\ref{fig:muhdep}) for color sextet. Thus, the theoretical prediction of $\rm NLL$ result for color sextet is inaccurate.

\begin{figure}[htb]
\centering
\subfigure[]{
	\includegraphics[width=0.42\textwidth]{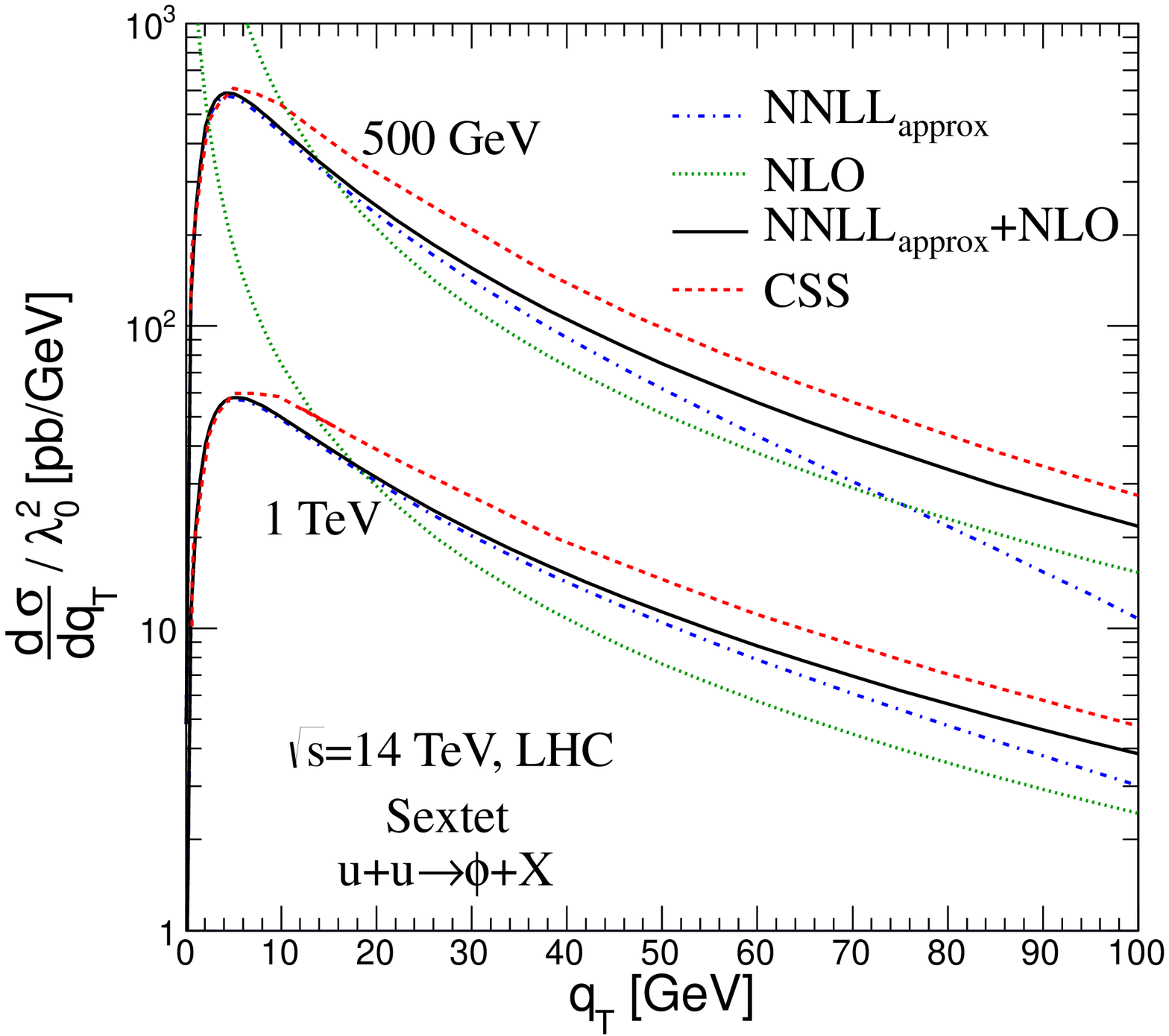}
	}
\subfigure[]{
	\includegraphics[width=0.42\textwidth]{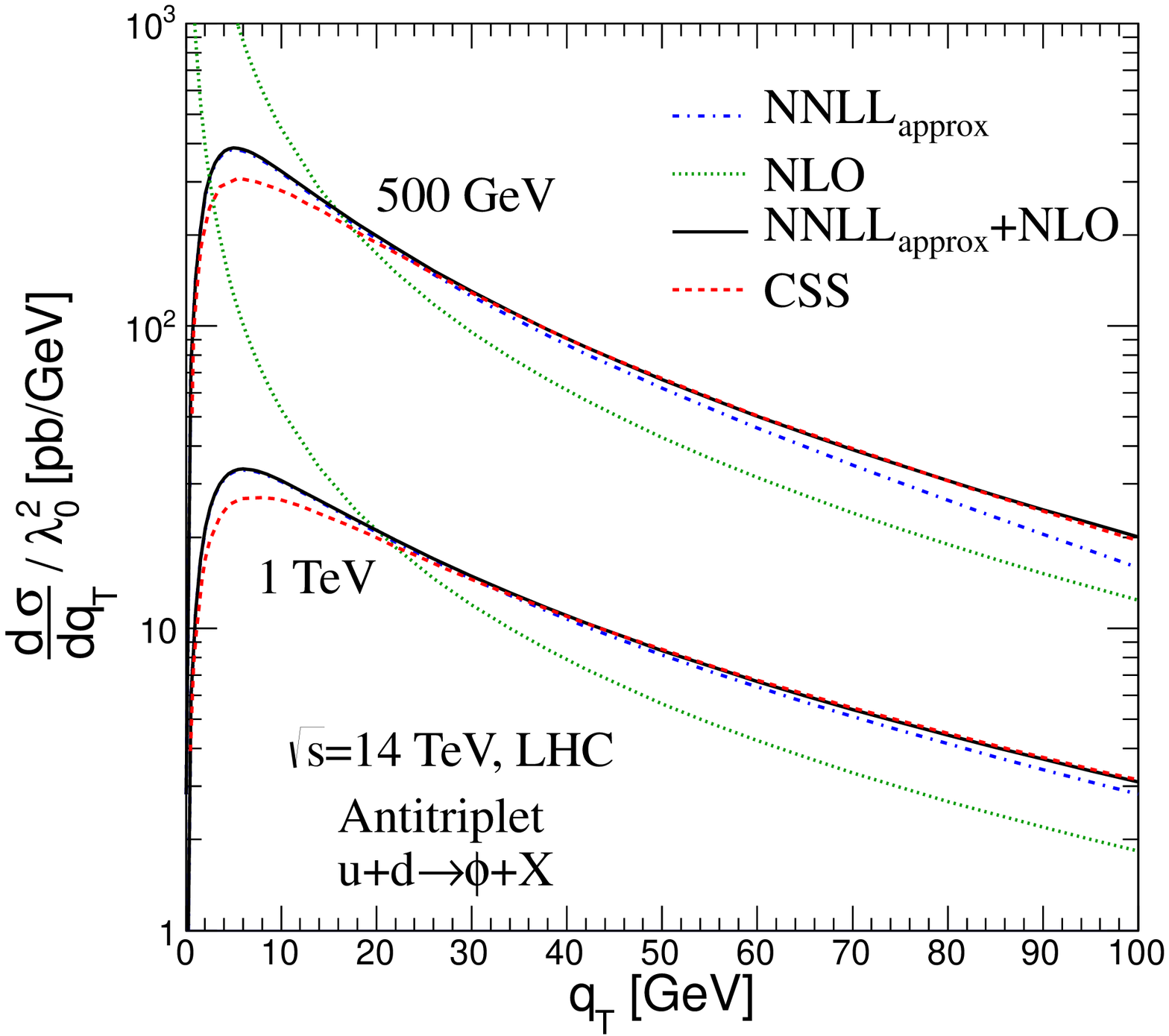}
	}
\caption{\label{fig:matched}
Transverse momentum distribution for the single colored scalar production with mass of $500\,{\rm GeV}$ and $1\,{\rm TeV}$ at the LHC. The results in the traditional framework developed by CSS are shown with red dashed lines.}
\end{figure}

Fig.~\ref{fig:matched} shows the transverse momentum distribution for both $m_\phi=500\,{\rm GeV}$ and $m_\phi=1\,{\rm TeV}$. Comparing with the work in Ref.~\cite{Han:2009ya}, our results have some differences. First, the peak of the $q_T$ distribution of $\rm NNLL_{approx}+NLO$ is suppressed by about 3\% for sextet and enhanced by about 25\% for antitriplet. Second, our low $q_T$ distributions peak around $4\sim 6.5\,\rm GeV$, while the peak region in Ref.~\cite{Han:2009ya} is around $5\sim 8\,\rm GeV$. These are due to the fact that our result of the resummation is presented at higher order than the one in Ref.~\cite{Han:2009ya}.

\begin{figure}[t!]
\begin{center}
\begin{tabular}{cc}
\includegraphics[width=0.42\textwidth]{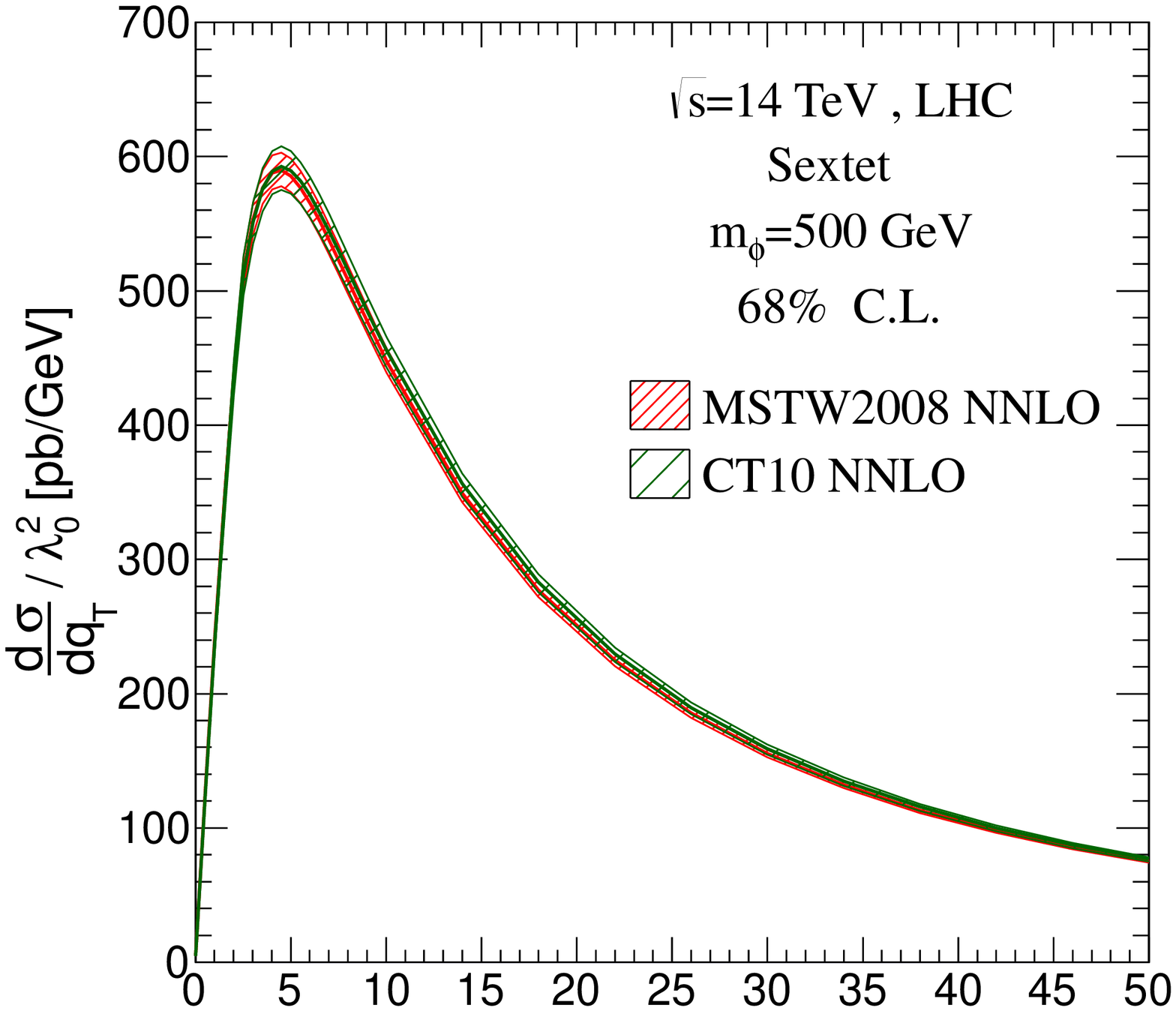}  &
\includegraphics[width=0.42\textwidth]{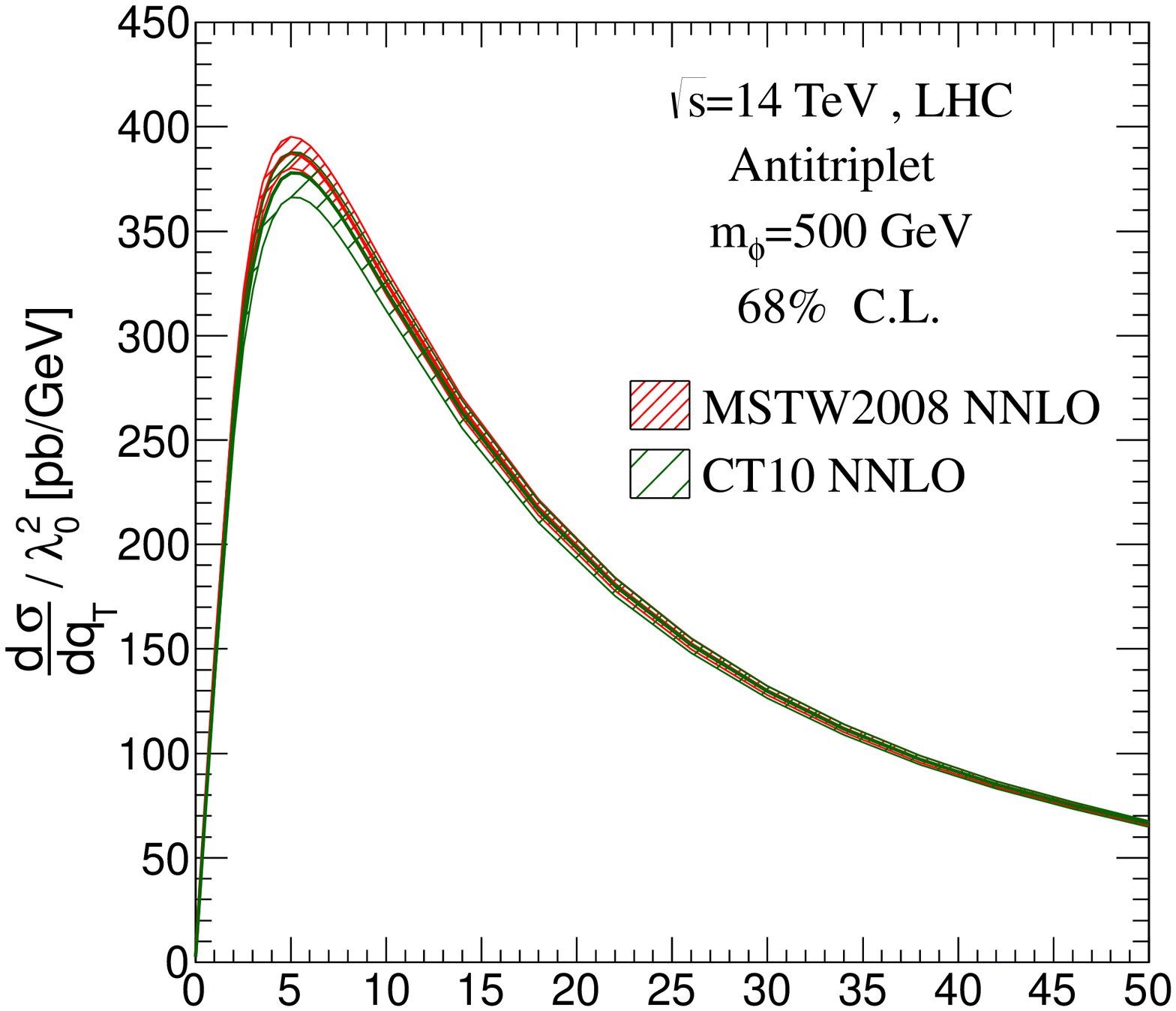}\\
\includegraphics[width=0.42\textwidth]{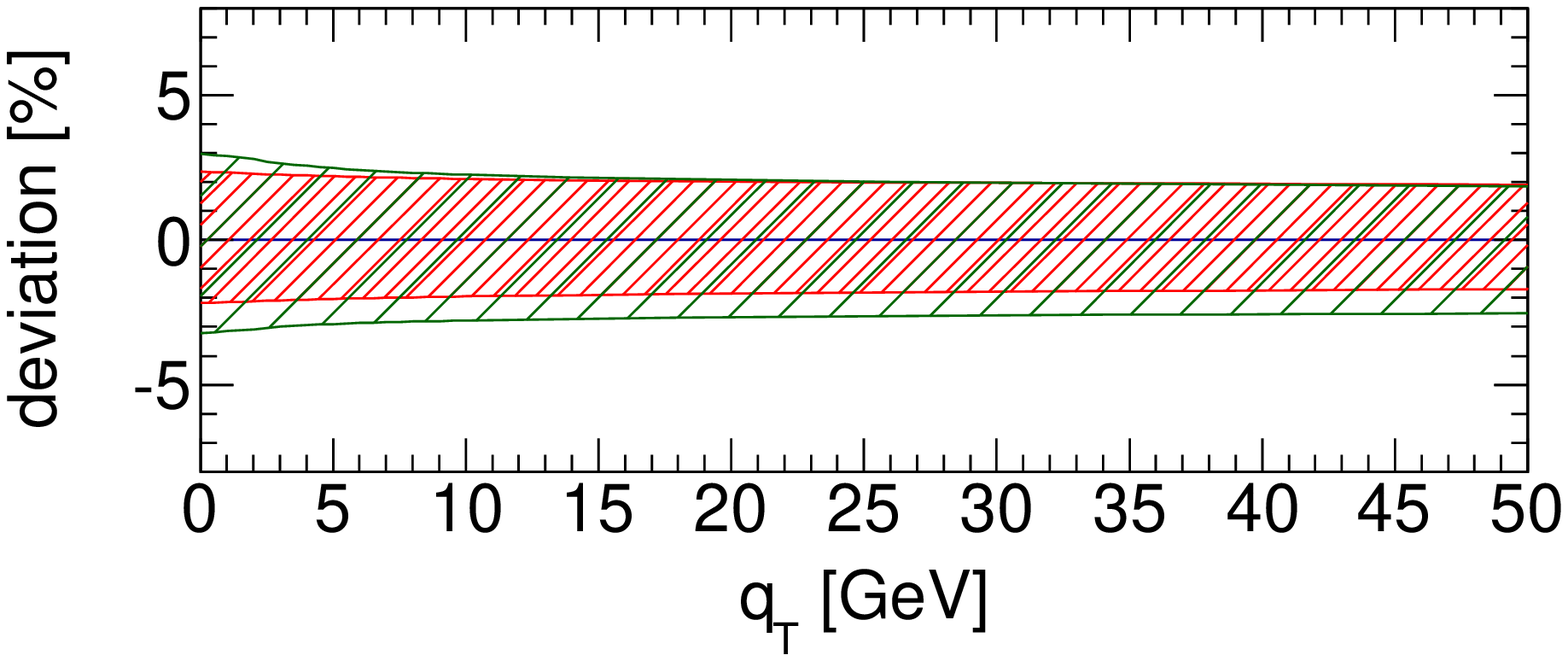}  &
\includegraphics[width=0.42\textwidth]{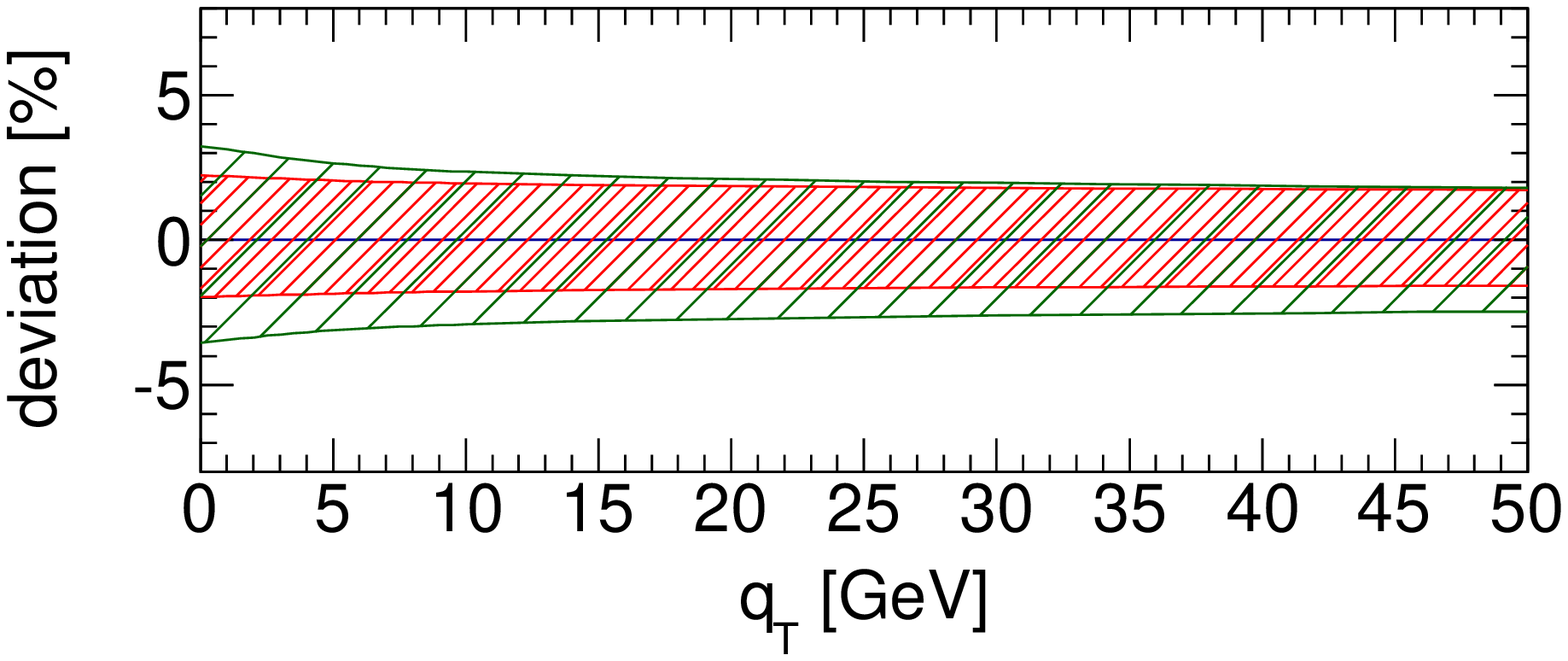}
\end{tabular}
\vspace{-0.5cm}
\end{center}
\caption{\label{fig:PDFsUC}
PDF uncertainties for the production of colored scalar particles at the LHC. The bands correspond to $1\sigma$ deviation variations of the PDF sets in MSTW2008 NNLO (red) and CT10 NNLO (green). At lower row we show the deviations to each central value.}
\end{figure}

In Fig.~\ref{fig:PDFsUC}, we show the PDF uncertainties of the $q_T$ distributions with MSTW2008NNLO \cite{Martin:2009bu} and CT10NNLO \cite{Lai:2010vv} PDF sets. For MSTW2008NNLO, the deviations are $\pm 2.5\%$ for $q_T\le 5\,\rm GeV$ and decrease to roughly $\pm 1\%$ near $q_T=50\,\rm GeV$, while for CT10NNLO, the PDF uncertainties are a little larger. The central values of differential cross sections with the two different PDF sets are almost identical to each other, with deviation smaller than $\pm 2.5\%$.
\section{Conclusion}\label{sec:conclusion}
We have studied the factorization and transverse momentum resummation effects in the single production of the color sextet (antitriplet) scalars at the LHC with the SCET. The soft function is calculated in analytic regularization at the NLO and its validity is demonstrated. From the comparison of the results for NLL and $\rm NNLL_{approx}$ resummation, we find the scale dependence is improved significantly in higher order. Comparing with the results in Ref.~\cite{Han:2009ya}, the peak of the $q_T$ distribution of $\rm NNLL_{approx}+NLO$ is suppressed by about 3\% for sextet and enhanced by about 25\% for antitriplet, respectively. In addition, our low $q_T$ distributions peak around $4\sim 6.5\,\rm GeV$, while the peak region in Ref.~\cite{Han:2009ya} is around $5\sim 8\,\rm GeV$. Also, we discuss the long-distance corrections to the transverse momentum spectrum, and show that they shift the peak positions about $0.2\,{\rm GeV}$ with $\Lambda_{\rm NP}=600\,{\rm MeV}$. Finally, we show that the PDF uncertainties are of order $\pm 2.5\%$ in the peak region.

\begin{acknowledgments}
We would like to thank Qing Hong Cao, Jian Wang and Ding Yu Shao for helpful discussions. This work was supported in part by the National Natural Science Foundation of China under Grants No. 11375013 and No. 11135003.
\end{acknowledgments}

\bibliography{pTResum}

\end{document}